\documentclass[groupadress,superscriptaddress,preprint,showpacs,preprintnumbers,amsmath,amssymb]{revtex4}

\usepackage{graphicx}
\usepackage{dcolumn}
\usepackage{natbib}
\usepackage{bm}
\newcommand{\BI}{BiMn$_2$O$_5$}
\newcommand{\Y}{YMn$_2$O$_5$}
\newcommand{\HO}{HoMn$_2$O$_5$}
\newcommand{\MNT}{Mn$^{3+}$}
\newcommand{\MNF}{Mn$^{4+}$}
\newcommand{\HOT}{Ho$^{3+}$}
\newcommand{\K}{$\overrightarrow{k}$}

\begin{document}

\preprint{APS/123-QED}

\title{Commensurate magnetic structures of $\mathbf{\emph{R}Mn_{2}O_{5}}$ (\emph{R}=Y, Ho, Bi) determined by Single-crystal Neutron Diffraction}

\author{C. Vecchini}
 \affiliation{ISIS Facility, Rutherford Appleton Laboratory, Chilton, Didcot, Oxfordshire, OX11 0QX, United
 Kingdom.}
 \affiliation{IESL - FORTH, P.O. Box 1527, Vassilika, Vouton, 711 10 Heraklion, Crete, Greece.}
\author{L. C. Chapon}
 \affiliation{ISIS Facility, Rutherford Appleton Laboratory, Chilton, Didcot, Oxfordshire, OX11 0QX, United Kingdom.}
\author{P. J. Brown}
 \affiliation{Institut Laue-Langevin, 6, rue Jules Horowitz, BP 156 - 38042 Grenoble Cedex 9 - France.}
\author{T. Chatterji}
\affiliation{Institut Laue-Langevin, 6, rue Jules Horowitz, BP 156 -
38042 Grenoble Cedex 9 - France.}
\author{S. Park}
 \affiliation{Rutgers Center for Emergent Materials and Department of Physics \& Astronomy, Rutgers University, Piscataway, New Jersey 08854, USA.}
\author{S-W. Cheong}
 \affiliation{Rutgers Center for Emergent Materials and Department of Physics \& Astronomy, Rutgers University, Piscataway, New Jersey 08854, USA.}
\author{P. G. Radaelli}
 \affiliation{ISIS Facility, Rutherford Appleton Laboratory, Chilton, Didcot, Oxfordshire, OX11 0QX, United Kingdom.}

\date{\today}

\begin{abstract}

Precise magnetic structures of \emph{R}$Mn_{2}O_{5}$, with \emph{R}
= Y, Ho, Bi in the commensurate/ferroelectric regime, have been
determined by single-crystal neutron diffraction. For each system, the
integrated intensities of a large number of independent magnetic Bragg reflections have been
measured, allowing unconstrained least-squares refinement of the
structures. The analysis confirms the previously reported magnetic
configuration in the ab-plane, in particular the existence of
zig-zag antiferromagnetic chains. For the Y and Ho compounds
additional weak magnetic components parallel to the c-axis were detected which are modulated in phase quadrature with the a-b components. This component
is extremely small in the \BI\ sample, therefore supporting symmetric
exchange as the principal mechanism inducing ferroelectricity.
For \HO, a magnetic ordering of the Ho moments was observed, which is consistent
with a super-exchange interaction through the oxygens.
For all three compounds, the point symmetry in the magnetically ordered state is m2m,
allowing the polar \textit{b}-axis found experimentally.

\end{abstract}

\pacs{25.40.Dn, 75.25.+z, 77.80.-e}
\maketitle

\section{\label{sec:level1}Introduction}

Materials in which there is strong interplay between ferroelectric and
magnetic ordering are attracting  current interest due to the
possibility of controlling the electric polarization by application
of a magnetic field or conversely the magnetisation with an electric
field. In the last few years, studies have been focussed on systems,
such as TbMnO$_3$ \cite{Kimura_Nature03,Hur_Nature04} and
\textit{R}Mn$_2$O$_5$\cite{Cheong_NatureMat}, for which the
ferroelectric transition coincides with the transition to a complex
antiferromagnetic ordered phase at low-temperature. These systems
are classified as \emph{improper} ferroelectrics, since the primary
order parameter is the magnetization, and their electrical
polarization is at least an order of magnitude lower than in proper
ferroelectrics\cite{ferroelectrics}. Nonetheless, the coupling
between the magnetization and the electrical polarization is strong, providing
an exciting playground in which to study the microscopic mechanisms that
govern the magneto-electric interaction. Moreover, the
aforementioned systems commonly present a high degree of magnetic
frustration, which seems closely linked to
the appearance of ferroelectricity\cite{Eerenstein_Nature}.\\
\indent The nature of the low-temperature state in these materials
has been investigated by a variety of experimental techniques
sensitive to small ionic displacements in a crystal (Raman, IR
spectroscopy)\cite{Garcia-Flores_JAP07, Aguilar_PRB06} and
scattering techniques sensitive to the magnetic order (Neutron,
resonant X-ray scattering)\cite{Noda_PhysB06, Han_JAP06, Lin_IEEE05,
Prokhnenko_PRL07, koo_2007}. An increasing number of theoretical
models have also appeared over the last two years that discuss the
origin of the magneto-electric coupling\cite{Khomskii_JMMM2006,
Yamasaki_PRL07, Cheong_NatureMat, Lawes_PRL05, ChaponPRL_Y}. Two
microscopic mechanisms have been put forward: on one hand the
antisymmetric \emph{inverse} Dzyaloshinskii-Moriya (DM) interaction,
requiring a non-collinear magnetic arrangement and on the other hand
exchange-striction due to symmetric exchange. In the DM model, the
polarization direction \textbf{P} (atomic displacement direction) is
given by the expression:
\begin{equation}
P \propto e_{12} \times (S_1 \times S_2)
\end{equation}
where S$_1$ and S$_2$ are neighboring spins and e$_{12}$ is the
propagation direction. The proportionality constant is related to
the strength of the spin-orbit coupling. It seems that this
mechanism explains the origin of ferroelectricity in many systems
recently investigated such as
TbMnO$_3$\cite{Yamasaki_PRL07,Lawes_PRL05, Cheong_NatureMat}.
Exchange-striction, which is intrinsically stronger that DM, does
not require non-collinearity, and has been proposed to explain the
appearance of ferroelectricity in RMn$_2$O$_5$\cite{ChaponPRL_Y} and
in the 'E' -phase of HoMnO$_3$\cite{Sergienko_PRL06}. Recent work by
Infra-Red absorption confirmed that both mechanisms described above
are relevant, as suggested by different polarization selection rules
for the electromagnons in \textit{R}MnO$_3$ and
\textit{R}Mn$_2$O$_5$\cite{Aguilar_PRB06}. It is important to note,
that due to the extremely small shifts of atoms in the
ferroelectric state with respect to their positions in the centro-symmetric
paraelectric phase, these displacement patterns have not yet been
properly determined. The mechanism which drives the ferroelectric
transition can therefore  often be more easily inferred from analysis
of the magnetic configuration derived from neutron diffraction experiments.\\
There has been continuing  interest in the complex magnetic structures
of \indent \textit{R}Mn$_{2}$O$_{5}$ with \textit{R} = Rare earth
(space group Pbam)  since the 1960s
\cite{Wilkinson_JPC1981, Gardner_JPC1988, Ratcliff_PRB05}. These
insulators order antiferromagnetically at low temperature (T$_N$
$\approx$ 40 K) with a propagation vector (k$_{x}$, 0, k$_{z}$) and
on further cooling undergo a series of magnetic transitions to both
commensurate and incommensurate magnetic phases. These materials
become ferroelectric at a temperature slightly below the N\'eel transition
at T$_N$ with values of electric polarization, ranging from 20 to
100 nC.cm$^{-2}$ depending on the \textit{R} ion, being largest in
the commensurate phase. The component of the magnetic propagation
vector along \textit{c} (k$_z$) depends on the size of the
\textit{R} cation, which suggests that the magnetic exchange along
the \textit{c}-direction is extremely sensitive to the interlayer
coupling\cite{Han_JAP06, MunozPRB_Bi}. Bi with r$_I$=131 pm
has the largest ionic radius of the series and in \BI\ k$_z$=1/2 whereas
 k$_z$=1/4 for \Y, \HO\ and TbMn$_2$O$_5$  with r$_I$=116,116,118 pm
respectively\footnote{ Ionic radius for a valence of 3+ and
8-coordinated configuration}. On the other hand, the onset and strength of
ferroelectricity does not seem to be dependent on r$_I$.\\
\indent The magnetic structures of this class of materials have mostly been
derived from neutron diffraction on polycrystalline samples.
Recently however, Noda and co-workers have reported results obtained by
single-crystal neutron diffraction for the commensurate magnetic
phase of \Y \cite{Noda_PhysB06}. Their analysis suggested the
presence of a small $c$-axis component of magnetic moment on the Mn
sites, which had not been detected
in previous powder measurements\cite{ChaponPRL_Y}. This
additional component is in phase quadrature with the major a-b
component and introduces a cycloidal modulation of the magnetic moments
which could, in principle, induce ferroelectric order through the
DM interaction. \\
\indent In the light of this recent work, we have
undertaken determination of the commensurate magnetic structures of three compounds
\BI, \Y\ and \HO\ using single-crystal neutron diffraction. Variation
of the cations allows a comparative study in which their influence on various
characteristics of the magnetic structures can be distinguished. The aim
of the study is twofold: first to search systematically for the
presence of the \textit{c}-axis modulation in several members of the series
so as to gauge its importance in promoting ferroelectricity.
Determination of the magnitude of this component for all three compounds
is especially important because the magnetic propagation vector  in \BI\
(k$_z$=$\frac{1}{2}$) differs
from that in \HO\ and \Y\ (k$_z$=$\frac{1}{4}$). Secondly, by studying \HO,
we can investigate the influence of a magnetic \textit{R} ion on
the arrangement of the Mn moments and determine the magnetic ordering of the Ho sublattice itself. Comparison
with \Y, which has the same propagation vector but a non-magnetic
\textit{R} site is of particular interest, since the values of the
electric polarization in the commensurate regime of the two systems
are significantly different.

\section{Experiment}

Single crystals of \emph{R}Mn$_{2}$O$_{5}$ (\emph{R}=Bi, Y, Ho) were
grown using B$_2$O$_3$/PbO/PbF$_2$ flux in a Pt crucible. The flux
was held at 1,280$^{\circ}$C for 15 hours and slowly cooled down to
950$^{\circ}$C at a rate of 1$^{\circ}$C per hour. Crystals grew in
the form of cubes. The samples used for the present work were
respectively of sizes: $\sim$ 4x4x4 mm$^3$ for \Y,  2x2x2 mm$^3$ for
\HO\ and 1x1x3 mm$^3$ for \BI, where the three dimensions refer to
the [110],
[$\overline{1}$10] and [001] directions. For the \HO\
crystal, additional faces  (100) and (010) were also visible. Single crystal diffraction
measurements were performed on the four-circle diffractometer D10 at
the ILL (Grenoble, France). Samples were checked for quality and
pre-aligned with the (001) direction oriented along the vertical
axis using the OrientExpress facility (ILL). Samples were mounted on
standard aluminum pins in a He exchange gas
cryostat. Data were collected with an incident neutron wavelength of
$\lambda$ = 2.36 \AA\ using an 80 mm$^2$ 2D micro-strip detector.
$\Omega$ scans around each Bragg reflection were performed, with
variable counting time depending on the intensity. Peak integration
was performed using the program \emph{Racer} (ILL), in two steps.
First, a library was built by fitting ellipsoidal shapes to a set
of strong reflections (I$>$3$\sigma$), these shapes were used in
a second pass to integrate all reflections. Data were
collected for all three samples in the ordered magnetic commensurate phases.
An additional dataset was
collected for the \Y\ sample in the low-temperature incommensurate
phase, but the results will be reported elsewhere. The data sets for \Y\
were collected at T = 25 K (nuclear and magnetic
scattering), those for \BI\ at T = 10 K (nuclear and magnetic scattering) and
for \HO\ at T = 50 K (nuclear scattering) and T = 25 K (magnetic
scattering). For each data-set, the list of integrated intensities
obtained were corrected for Lorentz-factors and normalized to an
arbitrary monitor count. Determination of both the
nuclear and magnetic structures were carried out using the FullProf
program suite\cite{Fullprof}. For each sample the following
procedure was applied: the published crystallographic models, space
group \textit{Pbam} were used to refine the nuclear data-sets.
Because a limited Q-range is accessible on D10, the crystallographic
parameters were fixed, with the exception of a global thermal
parameter to account for different temperatures. In addition, a
global scale factor was refined together with six parameters for
extinction correction, following the formulation of Becker-Coppens
for anisotropic extinction (extinction model number 4 in FullProf).
This correction is particularly important for  \Y\ because of the large
size of the crystal. The scale factors obtained by refining
the nuclear structure were used and fixed for refinement of the magnetic structures.\\

\section{Results}
\subsection{Crystal structure and formalism}
\begin{figure*}[h!]
\includegraphics[scale=0.20, angle=0]{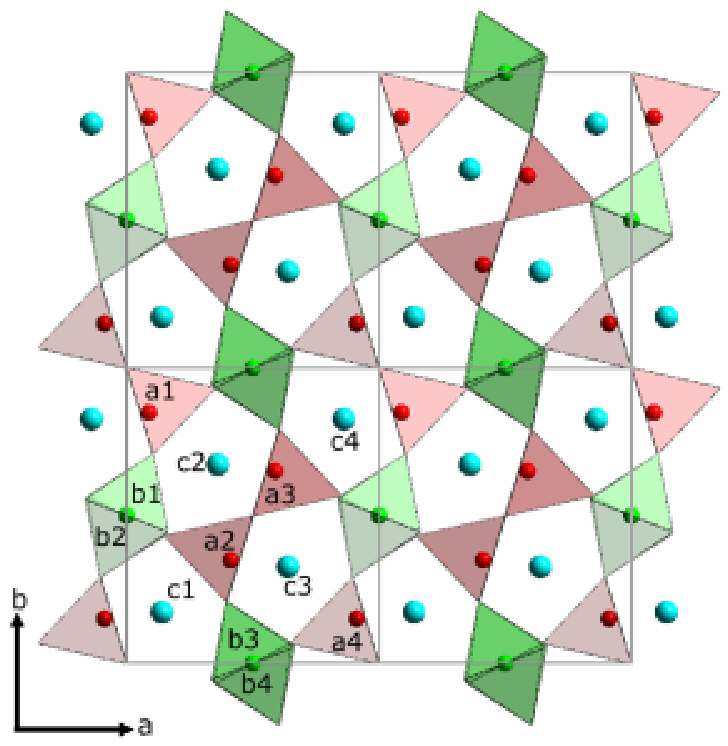}
\includegraphics[scale=0.20, angle=0]{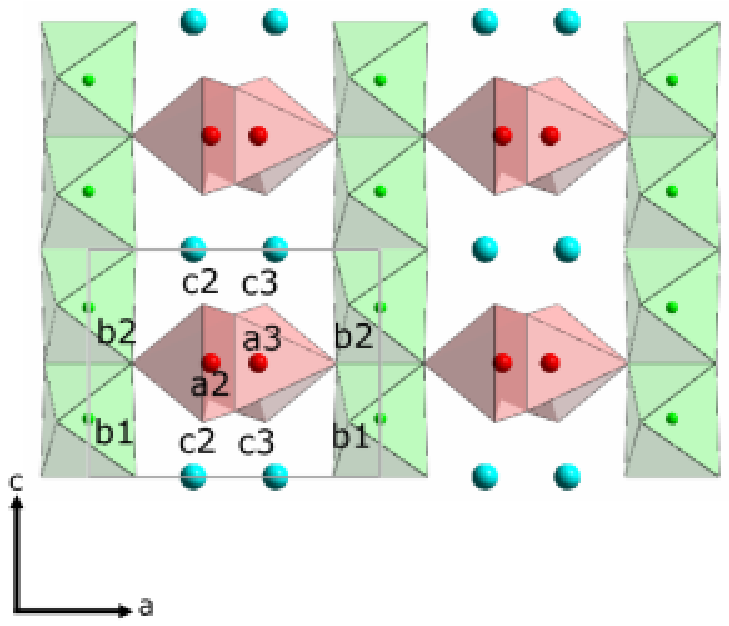}
\caption{(Color Online) Crystal structure of $RMn_{2}O_{5}$
projected in the \emph{ab}-plane (left panel) and \emph{ac}-plane
(right panel). Green, red and blue spheres correspond to $Mn^{4+}$,
$Mn^{3+}$ and $R^{3+}$ ions, respectively. Mn-O polyhedra are shown
with the same color scheme. The thin black line represent the
crystallographic unit-cell. Correspondence between the labels and
atomic positions are described in Table \ref{HoMn2O5Fixedtable}. }
\label{y_nuc}
\end{figure*}

\indent The crystal structure and topology of the magnetic
interactions have been described in detail in
\cite{ChaponPRL_Y,BlakePRB}. The different atomic positions for
\MNT, \MNF\ (and \HOT ) have been numbered according to reference
\cite{BlakePRB}. To facilitate the identification of all magnetic
sites in the figures, additional labels have been included: a for
\MNT, b for \MNF (c for \HOT). The crystal structure of
\textit{R}Mn$_2$O$_5$ (isostructural throughout the series) is
displayed in Fig. \ref{y_nuc} and the correspondence between labels and
atomic positions can be found in table \ref{YMn2O5Fixedtable}. \MNT\
(a) is five-coordinated by oxygen ions, in square-pyramid geometry,
while \MNF\ (b) is six-coordinated in distorted octahedral geometry.
There are four symmetry equivalent \MNT\ sites and four \MNF\ sites
per unit-cell. The pyramids centered at a2 and a3 (also a1 and a4)
share an edge of their basal plane. All \MNT\ are located on a
mirror plane at z=0.5. The octahedra
coordinating the \MNF\
sites b1 and b2 (also b3 and b4) share edges to form chains
running along the \textit{c}-direction. The b1(b3) atoms with  z$\sim$0.25
and b2(b4) with(1-z)$\sim$0.75 lie below and above the \MNT\ layer. It
should be noted that the sites b$_1$ and b$_2$ also  b$_3$ and b$_4$ are
related by a center of symmetry. The \MNF\ octahedra share corners
with \MNT O$_5$ pyramids to form layers in the \textit{ab}-plane.
The \textit{R} ions (labeled c for \HO) occupy the (x,y,0) site, of
multiplicity four, and are all positioned in the z=0 layer
separating two \MNT /
\MNF\ layers. \\
\indent The magnetic structures have been analyzed using the
propagation-vector formalism, in which magnetic moments for all
sites described above are expanded in Fourier series. We briefly
describe the conventions used in the next paragraph. For a magnetic
structure with a single propagation vector \K, which is the
relevant case in this study, a magnetic moment of atom type j in the
crystallographic unit-cell l ($\overrightarrow{m_{lj}}$) is written:
\begin{equation}
\overrightarrow{m_{lj}}=\sum_k
\overrightarrow{S_{kj}}\cdot e^{-2\pi\cdot i\cdot(\overrightarrow{k}\cdot \overrightarrow{R_l}+\phi_j)}
\end{equation}
where the sum runs over \K\ and -\K\, if these vectors are not
related by a reciprocal lattice vector (case of \Y\ and \HO) or
uniquely \K\ otherwise (case of \BI). $\overrightarrow{R_l}$ is a
pure lattice translation and $\phi_j$ is a phase factor. The
magnetic structure is fully determined by giving a set of Fourier
coefficients and phases ($\overrightarrow{S_{kj}}$,$\phi_j$) for
each j, i.e seven components per site in the most general case since
the Fourier coefficients are complex quantities. We note that
because $\overrightarrow{m_{lj}}$ is a real quantity, the constraint
$\overrightarrow{S_{-kj}}$=$\overrightarrow{S_{kj}^*}$ (where *
denotes the complex conjugate) is required. It is important to
mention that $\phi_j$ are relative phases and are equal for sites
belonging to the same orbit, i.e. sites that can be transformed into
each other by an operation of the magnetic little group. It is
different than a global phase, that can't be determined by
diffraction, which rephase the entire magnetic structure, i.e. apply
simultaneously to all magnetic sites. Additional constraints exist
between ($\overrightarrow{S_{kj}}$,$\phi_j$) of symmetry-related
sites. These constraints are determined by representation analysis
and will be briefly described in the
following sections, based on previously published work\cite{radaelli-2006}.\\

\subsection{YMn$_{2}$O$_{5}$}
\subsubsection{Magnetic structure}
\begin{figure*}[h!]
\includegraphics[scale=0.20, angle=0]{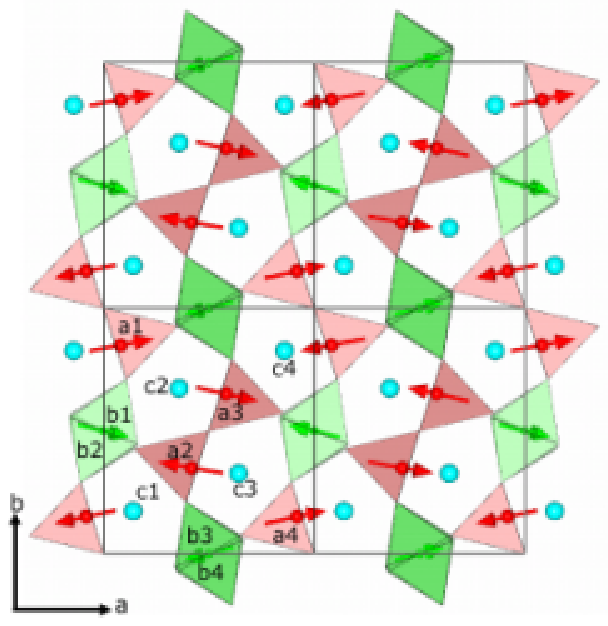}
\caption{(Color Online) Magnetic structure of $YMn_{2}O_{5}$
projected in the \textit{ab} plane. The structure is shown in two
unit-cells, marked by thin black lines, along the \textit{a}- and
\textit{b}-axis. The green and red arrows represent magnetic moments
on $Mn^{4+}$ and $Mn^{3+}$ sites respectively. Corresponding Mn-O
polyhedra are shown with the same colors. Blue spheres represent Y
ions. The different sites are labeled according to Table
\ref{YMn2O5Fixedtable}.}
\label{y_ab}
\end{figure*}

\begin{figure*}[h!]
\includegraphics[scale=0.20, angle=-0.7]{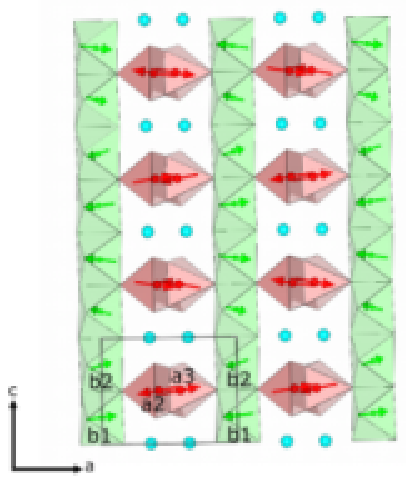}
\includegraphics[scale=0.20, angle=-0.4]{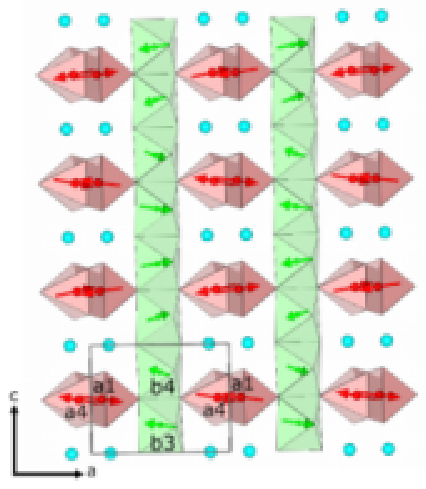}
\caption{(Color Online) Magnetic structure of $YMn_{2}O_{5}$
projected in the \textit{ac}-plane shown within four unit-cells along
\textit{c} and two unit-cells along \textit{a}. The projection is
shown separately for magnetic sites belonging to the first AFM chain
(left) and the second chain (right) (See text for details). The
green and red arrows represent magnetic moments on $Mn^{4+}$ and
$Mn^{3+}$ sites respectively. Corresponding Mn-O polyhedra are shown
with the same colors. Blue spheres represent Y ions. The different
sites are labeled according to Table \ref{YMn2O5Fixedtable}.}
\label{y_ac}
\end{figure*}

\begin{figure*}[h!]
\includegraphics[scale=0.35, angle=0]{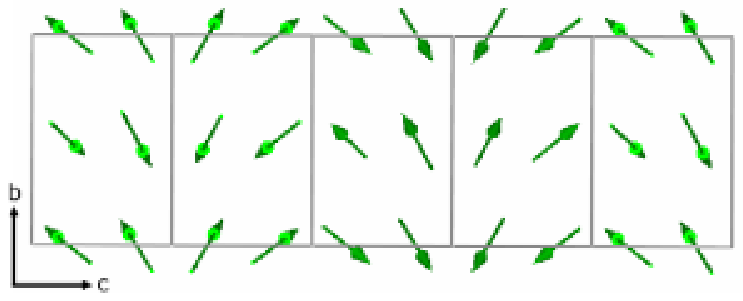}
\caption{Magnetic structure of \Y\ projected in the
\textit{bc}-plane and showing only the Mn$^{4+}$ moments. The figure
shows the small helicoidal modulation generated by the out-of-phase
\textit{c}-component. To highlight this weak modulation, the moment
have been scaled by a factor five with respect to what represented
in the other Figures. }
\label{y_bc}
\end{figure*}

The propagation vector for the Y-compound in the commensurate phase
(CM) is $\overrightarrow{k}$=(1/2, 0, 1/4). Symmetry analysis
\cite{radaelli-2006} indicates that the \MNF\ sites is split into
two orbits because the \textit{m$_z$} mirror-plane operation does
not belong to the little group. The first and second orbits contain
respectively the b1/b3 sites and b2/b4 sites. The $\phi_j$ phases
for each orbit can therefore be non-equal and treated as free
parameters during the refinement. On the other hand, the phases of
the \MNT\ sites, belonging to a single orbit, must be equal. We
arbitrarily chose to fix the phase of the \MNT\ sites at 0.125 (one
phase must be fixed since structure factors do not depend on a
global phase), for comparison with results reported earlier from
powder data \cite{ChaponPRL_Y} describing a configuration with equal
moments. In this case, representation analysis does not impose any
constraints on the magnitude and direction of the Fourier
coefficients for both \MNT\ and \MNF\ sites\cite{radaelli-2006}. For
each site, six independent Fourier coefficients are in principle
refinable. The unconstrained model has therefore fifty parameters
and describe the most general magnetic structure. For typical
magnetic structures described by either spin-density waves (SDW) or
simple cycloidal modulations, some of these parameters will be fully
correlated. Refining an unconstrained model against the 355
independent magnetic reflections fails due to the large number of
parameters and the presence of strong correlations. In order to find
possible models, a global optimization algorithm, using the
simulated annealing procedure described in \cite{Juan_Nature}, was
employed. Systematically, the Fourier coefficients for all magnetic
sites were found to be principally along the \textit{a}-axis,
whether real or imaginary. In addition, a weaker \textit{b}-axis
component was found with the same character (real or imaginary) as
the \textit{a}-component. These two components define a vector, say
v$_1$ in the \textit{ab}-plane.
A small \textit{c}-axis component
 in phase quadrature with the \textit{a}
and \textit{b}-components was also identified. The structure
therefore has a small helicoidal modulation, with rotation axis given by
the cross product of v$_1$ with a vector along z. A model with this
 configuration, i.e one in which the Fourier coefficients for all
magnetic sites are real in the \textit{ab}-plane
and imaginary along \textit{c}, has been constructed, it has 26
refinable parameters. The lower number of free parameters allowed
a direct least-square refinement to be made which converged after a
few cycles. The structure factors calculated with this
model are shown, plotted against those observed, in Fig. \ref{refinements}.
The model was found to reproduce the observed data very well with a
Magnetic R(F) factor\footnote{R(F) factor is defined as: $R(F)=100
\{\sum_{\bf{h}}
|F_{obs,\bf{h}}-F_{calc,\bf{h}}|\}/\{\sum_{\bf{h}}{|F_{obs,\bf{h}}|}\}$,
where $F_{obs}$ and $F_{calc}$ are respectively the observed and
calculated structure factors for a given reflection $\bf{h}$} of
4.57\% (Fig. \ref{refinements}). For this particular sample, the
large value of $\chi^{2}$=163 is explained by high statistics of the
data obtained on a large sample with long counting time. The refined
values of the Fourier coefficients are displayed in Table
\ref{YMn2O5Fixedtable} and the magnetic structure projected along
two different crystallographic directions in Fig. \ref{y_ab} and
Fig. \ref{y_ac}. The magnetic arrangement in the \textit{ab}-plane
(Fig. \ref{y_ab}) is equivalent to that derived in our previous work
from powder data\cite{ChaponPRL_Y}. The moment directions are within
$\sim$10$^{\circ}$ of the \textit{a}-axis for \MNT and
$\sim$14$^{\circ}$ for \MNF, defining zig-zag antiferromagnetic
(AFM) chains running along \textit{a}. One of these chains
links b1, b2, a2 and a3, the other  b3, b4, a4 and a1. In
addition, there is a significant \textit{c}-component on all magnetic
sites as suggested by Noda et al.\cite{Noda_PhysB06}, which was not
found in our earlier analysis of powder data. The
\textit{c}-components have almost the same magnitude as the
\textit{b}-components on both Mn sites. Their imaginary character
means that the modulation is in quadrature with the real
\emph{in-plane} components. This can be seen in Fig. \ref{y_ac};
for example for the a2 site the  in-plane components of
the magnetic moment follow a (- + + -) sequence across four
unit-cells along \textit{c} whilst the c-component follows a (- - + +)
sequence. This result is also in
agreement with the work of Noda et al.\\
The refined values of the phases for sites b1/b3 and b2/b4,
respectively 0.095(2) and 0.156(2), are slightly different from the
fixed values of 0.125 used for fitting powder data. Consequently,
and in addition to the existence of a \textit{c}- component, the
magnetic moments on site b1 and b2 (as well as b3 and b4) are of
different magnitudes as shown in Fig \ref{y_ac}. We also note that
there is a symmetric deviation of these phases with respect to \MNT\
phase of 0.125 ( (0.125-$\delta$) for the first orbit and
(0.125+$\delta$) for the second orbit). The identification of a
small \textit{c}-component has direct implications on the nature of
the magnetic configuration, previously described as a pure SDW:
small cycloidal modulations propagating along the edge-sharing
Mn$^{4+}$O$_6$ chains are observed in the \textit{ac} and
\textit{bc} -planes (Fig. \ref{y_bc}). In the \textit{ac}-plane, the
moment rotation direction is clockwise for \MNF\ chains formed by
the b1 and b2 sites (left panel of Fig. \ref{y_ac}) and
anti-clockwise for chains formed by the b3 and b4 sites (right panel
of Fig. \ref{y_ac}). In the \textit{bc}-plane, the weak cycloidal
modulation, highlighted in Fig. \ref{y_bc}, rotates in the same
direction along b1/b2 chains and b3/b4 chains. The \MNT\ moments are
also slightly tilted along the \textit{c}-direction, remaining
almost collinear to adjacent \MNF\ magnetic moments in a given AFM
chain. Therefore, the presence of antiferromagnetic chains running
along the \textit{a}-axis, identified earlier as an important
characteristic of these systems, remain valid despite the small
\textit{c}-component.

\subsubsection{Domains}

\begin{figure*}[h!]
\includegraphics[scale=0.20, angle=0]{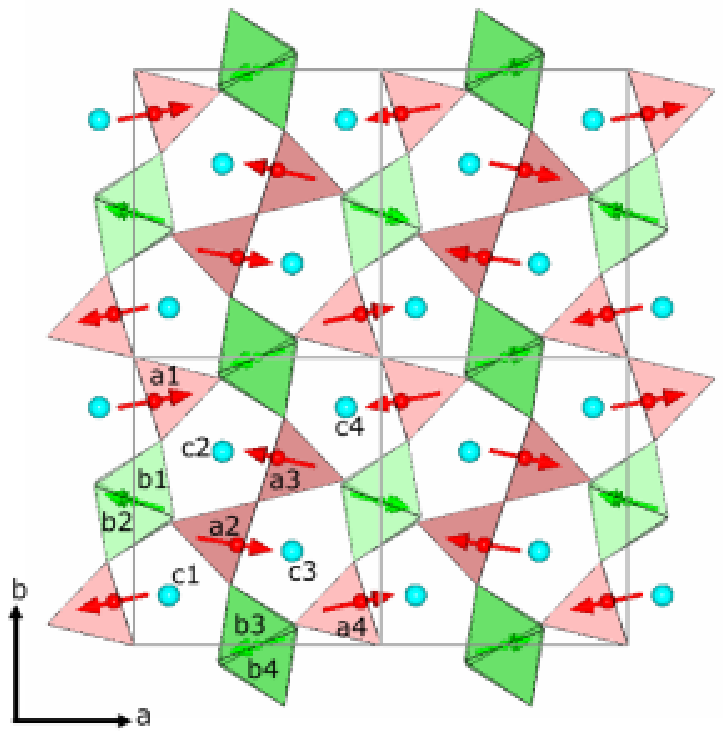}
\includegraphics[scale=0.18, angle=0]{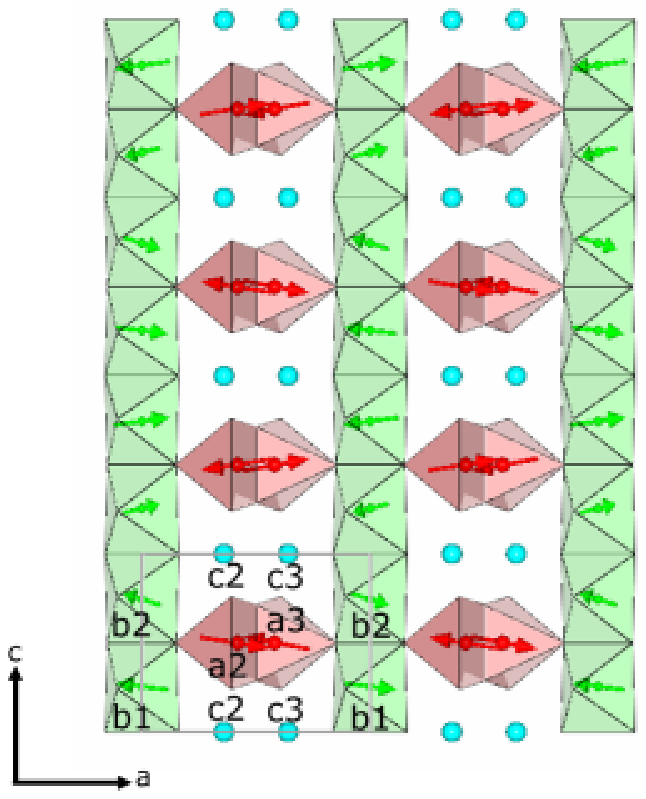}
\caption{(Color Online) Magnetic structure of $YMn_{2}O_{5}$
projected in the \emph{ab}-plane (left panel) and the \emph{ac}-plane (right panel) for the second domain, obtained by applying
inversion symmetry on the magnetic structure of the first domain
(Fig. \ref{y_ab}, \ref{y_ac})}
\label{y_ab_d2}
\end{figure*}

\indent The arm of the star of k is made only of the two vectors
$\overrightarrow{k}$=($\frac{1}{2}$,0,$\frac{1}{4}$) and
-$\overrightarrow{k}$ since the vectors
(-$\frac{1}{2}$,0,$\frac{1}{4}$) and
($\frac{1}{2}$,0,-$\frac{1}{4}$) are related to k and -k by a
reciprocal lattice vector. Therefore, a single k-domain contributes
to the scattering. On the other hand, one must take into account the
"orientation" domains, i.e. domains that are obtained by applying
symmetry operations of the paramagnetic group that are not valid
operations (lost) of the magnetic structure. Four symmetry
operations of the paramagnetic group (combined or not with complex
conjugation) leaves the magnetic structure unchanged, as shown in
Table \ref{SummetryTable}. We refer to this domain as domain one
(D1). The other four symmetry operations all transform the magnetic
structure described in the previous section into an inequivalent one
(domain 2, D2, shown in Fig. \ref{y_ab_d2}). No other domains exist,
and D1 and D2 are simply related by inversion symmetry. Inversion
symmetry operates differently on different magnetic sites: for
b$_3$, b$_4$, a$_1$ and a$_4$, belonging to one of the zig-zag AFM
chain, the application of inversion symmetry leaves the a and b
moment components unchanged while it changes the sign of the
c-component. In contrast for sites b$_1$, b$_2$, a$_2$ and a$_3$
(the other AFM chain), inversion symmetry reverses the sign of the a
and b components while preserving the c component. Since D1 and D2
are related by inversion symmetry, the diffracted intensities
arising from both domains are identical for a non-polarized
diffraction experiment and refining
the data assuming any domain leads to the same result. We note that the two domains are also related by a simple rotation by 180$^{\circ}$ along the \textit{z}-axis or the \textit{x}-axis.\\
\indent From the list of symmetry operations (Table.
\ref{SummetryTable}) that leave the magnetic structure invariant
(modulo complex conjugation operation), one derives the point
group $m2m$ in the magnetically ordered phase, in agreement with
that obtained earlier from co-representation
analysis\cite{radaelli-2006}. This point group is compatible
with ferroelectricity along the \textit{b}-axis, irrespective of
the microscopic magneto-electric mechanism.

\subsection{HoMn$_{2}$O$_{5}$}
\begin{figure*}[h!]
\includegraphics[scale=0.20, angle=0]{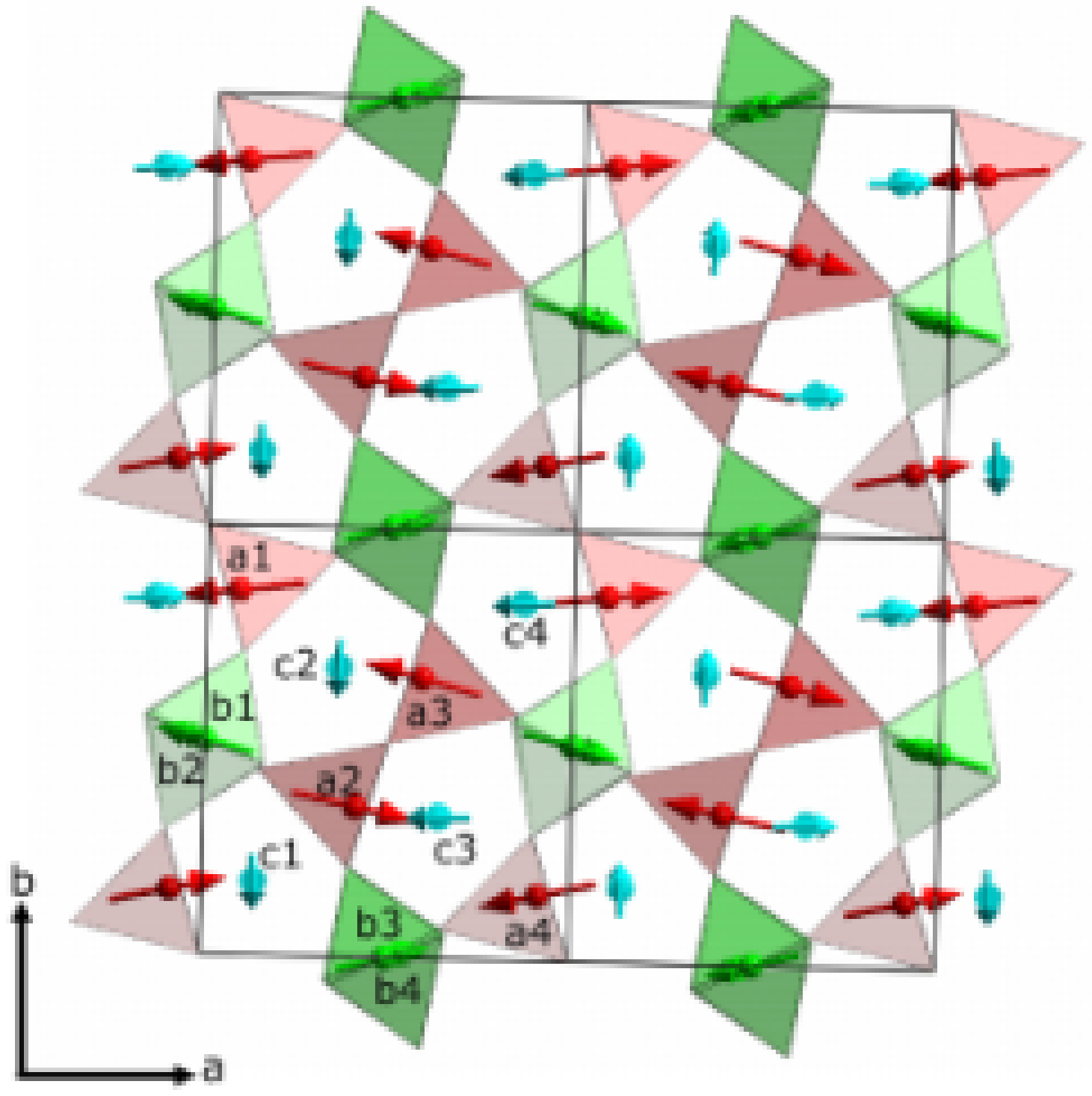}
\includegraphics[scale=0.20, angle=0]{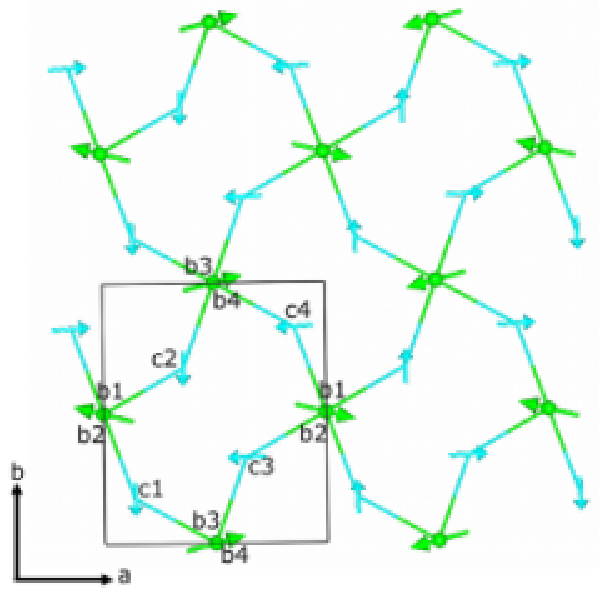}
\caption{(Color Online) Magnetic structure of $HoMn_{2}O_{5}$
projected in the \emph{ab} plane. The structure is shown in two
unit-cells, marked by thin black lines, along the \textit{a}- and
\textit{b}-axis. The green, red and blue arrows (left panel) represent magnetic moments
on $Mn^{4+}$, $Mn^{3+}$ and $Ho^{3+}$ sites respectively. Corresponding Mn-O
polyhedra are shown with the same colors. For clarity, in the right panel only $Mn^{4+}$ and $Ho^{3+}$ (belonging to an ordered $Ho$ layer) magnetic moments are shown respectively in green and blue color. Lines connecting $Mn^{4+}$ and $Ho^{3+}$ represent interactions (See text for details). The different sites are labeled according to Table \ref{HoMn2O5Fixedtable}.}
\label{Ho_ab}
\end{figure*}

\begin{figure*}[h!]
\includegraphics[scale=0.20, angle=0.0]{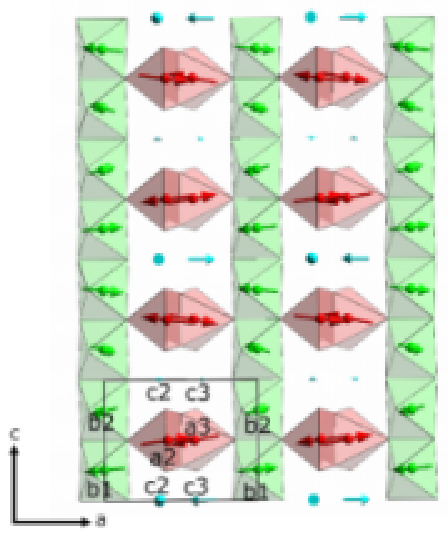}
\includegraphics[scale=0.20, angle=0.]{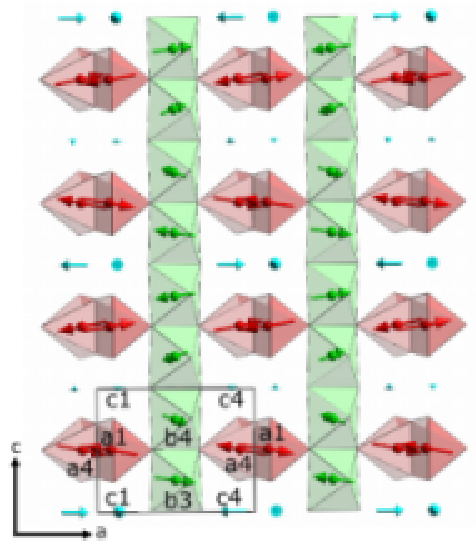}
\caption{(Color Online) Magnetic structure of $HoMn_{2}O_{5}$
projected in the \emph{ac} plane shown within four unit-cells along
\textit{c} and two unit-cells along \textit{a}. The projection is
shown separately for magnetic sites belonging to the first AFM chain
(left) and the second chain (right) (See text for details). The
green, red and blue arrows represent magnetic moments on $Mn^{4+}$, $Mn^{3+}$ and $Ho^{3+}$ sites respectively. Corresponding Mn-O polyhedra are shown
with the same colors. The different
sites are labeled according to Table \ref{HoMn2O5Fixedtable}.}
\label{Ho_ac}
\end{figure*}

\HO\ orders AFM at T = 44 K first with an incommensurate propagation
vector $\overrightarrow{k}$=(0.48, 0, 0.245), followed by a lock-in
at T = 38 K to a commensurate $\overrightarrow{k}$=($\frac{1}{2}$,
0, $\frac{1}{4}$). The magnetic structure has been determined in the
commensurate phase from 381 independent magnetic reflections, using
the same procedure as that described in the case of \Y. However in
this case, the magnetic ordering of Ho is an additional complexity
which increases the number of free parameters. Attempting to refine
the data without a magnetic contribution on the Ho sites, leads to
poor agreement factors (R(F)=20.2\%) as already reported from
analyzing powder diffraction patterns\cite{BlakePRB}. Preliminary
measurements with resonant X-ray scattering at the L$_{III}$ edge
directly confirm the magnetic ordering of Ho. Symmetry analysis
indicates that the four equivalent Ho positions belong to the same
orbit and therefore only one phase parameter is required. However,
there are no constraints between components of the moment of
different sites, which were refined independently. The final
refinement contained 45 parameters and was found to reproduce the
data very well with structure factor R(F)=3.96\% and $\chi^2$=5.82.
The magnetic structure is similar to the \Y\ one, with the
appearance of almost collinear
...-Mn$^{4+}$-Mn$^{3+}$-Mn$^{3+}$-Mn$^{4+}$-... AFM zig-zag chains
in the \emph{ab}-plane. These chains make an angle of
$\sim$14$^{\circ}$ with respect to the a-axis. We again found a small
\textit{c}-component (Fig. \ref{Ho_ac}) of moment on all Mn
magnetic sites, in quadrature with the in-plane components and
of magnitude similar to that observed for \Y. The refined phase for
the Ho sites is shifted by 0.126(2) with respect to the Mn$^{3+}$
sites. The Ho moments at the c3/c4 sites are parallel to
the \textit{a}-axis and those at the c1/c2
sites to the \textit{b}-axis (Fig. \ref{Ho_ab} and Table. \ref{HoMn2O5Fixedtable}). It
is possible that a smaller, $\pi/2$ out of phase component along \textit{c}
may exist, but the refined magnitudes are only three times larger than the error
bars. Although the magnetic configuration of the  Ho sites is identical
to that derived
earlier by simulated annealing using powder data\cite{BlakePRB},
the orientation of the Ho moments in the
\textit{ab}-plane is quite different.  The values of the Ho moments,
ranging from 1.1 to 1.3
$\mu_B$, are much smaller than the fully saturated value
for a J=8 state. This can be partly explained by splitting of
the fully degenerate J=8 state by the crystal field, but typical
crystal-field energies in oxides will lead to a moment much larger
than observed here\cite{Amoretti_CF94}. Additional reduction can be due
thermal excitation, at 25K, of levels above
the ground state (Ho$^{3+}$ is a non-Kramer ion characterized
by singlet levels).
Moreover, dipolar field calculations show no direct
correlation between the local field generated by neighbors magnetic
Mn sites and the magnitude and direction of the Ho moments.
Therefore we argue that the Ho moment of $\sim$1.1$\mu_B$ in \HO\ is
mainly due to super-exchange interactions via the 2p oxygen
orbitals. Exactly half of the Ho layers show ordered magnetic
moments in the \textit{ab}-plane, while the other half show only a
small moment along the \textit{c}-direction (Fig. \ref{Ho_ac}).
Accordingly, in the former type of layers, we can clearly show that
the directions of the moment are systematically arranged along the
resultant of AFM super-exchange interactions with the four
nearest-neighbor Mn$^{4+}$ sites (Fig. \ref{Ho_ab} and Fig.
\ref{Ho_ac}). The net moment on each Ho sites is pointing either
along the \textit{a}-axis (sites c3 and c4) or \textit{b}-axis
(sites c1 and c2), resulting from the cancelation of respectively
the \textit{b}- and \textit{a}-components of neighbor Mn$^{4+}$
moments. Also, the resultant of the super-exchange is larger for
sites c3 and c4 (along the \textit{a}-direction) than for sites c1
and c2 (along \textit{b}), which should lead to different Ho moments
on these sites, as we appear to observe experimentally (Table
\ref{HoMn2O5Fixedtable}). A more precise determination of the
respective magnitude of the moments on sites c1/c2 and c3/c4 is
currently underway by resonant X-ray scattering. Calculations show
that azimuthal scans on magnetic Bragg peaks at the Ho L$_{III}$
edge are extremely sensitive to these parameters (as well as being
Ho selective).\\
\indent The magnetic configuration of the Ho sublattice also breaks
inversion symmetry since moments on sites c2 and c3 (respectively c1
and c4) are pointing in different directions. As in the case of \Y,
there are only two "orientation" domains to consider, related by
inversion symmetry and therefore contributing equally to the
diffracted intensities. The symmetry operations and corresponding
domains listed in Table \ref{SummetryTable} are also valid for \HO.
The magnetic structure of domain 2 including the transformed Ho
moments (not shown) is easily derived from domain 1 by applying
inversion symmetry. In this case, the magnetic point group is also
m2m, which supports ferroelectricity along the \textit{b}-axis.

\subsection{BiMn$_{2}$O$_{5}$}
\begin{figure*}[h!]
\includegraphics[scale=0.25, angle=0]{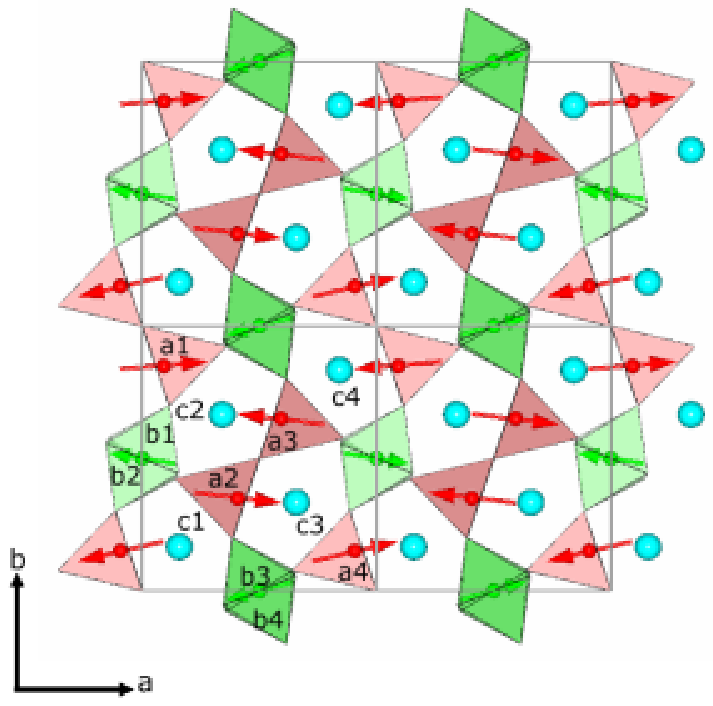}
\caption{(Color Online) Magnetic structure of $BiMn_{2}O_{5}$
projected in the \emph{ab} plane. The structure is shown in two
unit-cells, marked by thin black lines, along the \textit{a}- and
\textit{b}-axis. The green and red arrows represent magnetic moments
on $Mn^{4+}$ and $Mn^{3+}$ sites respectively. Corresponding Mn-O
polyhedra are shown with the same colors. Blue sphere represent Bi
ions. The different sites are labeled according to Table
\ref{BiMn2O5table}.}
\label{Bi_ab}
\end{figure*}

\begin{figure*}[h!]
\includegraphics[scale=0.20, angle=0.0]{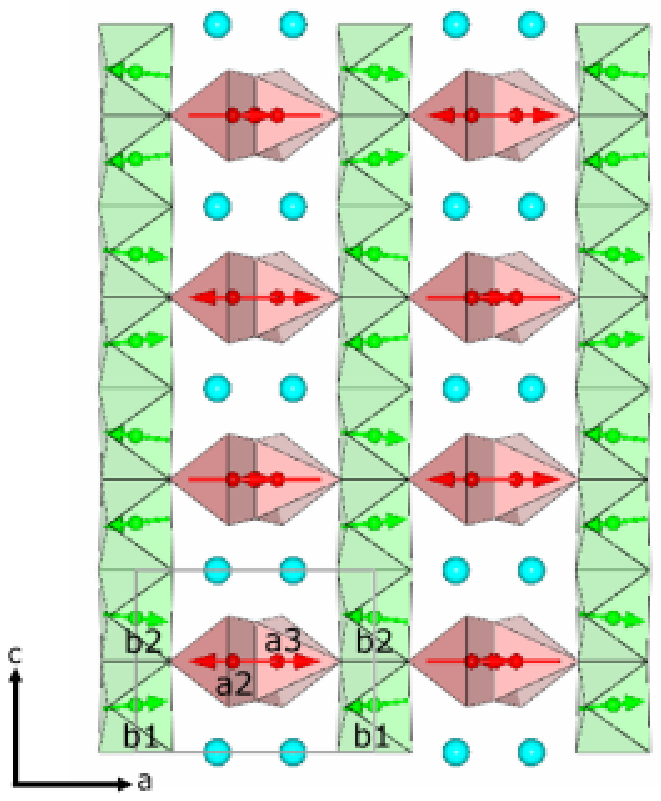}
\includegraphics[scale=0.20, angle=0]{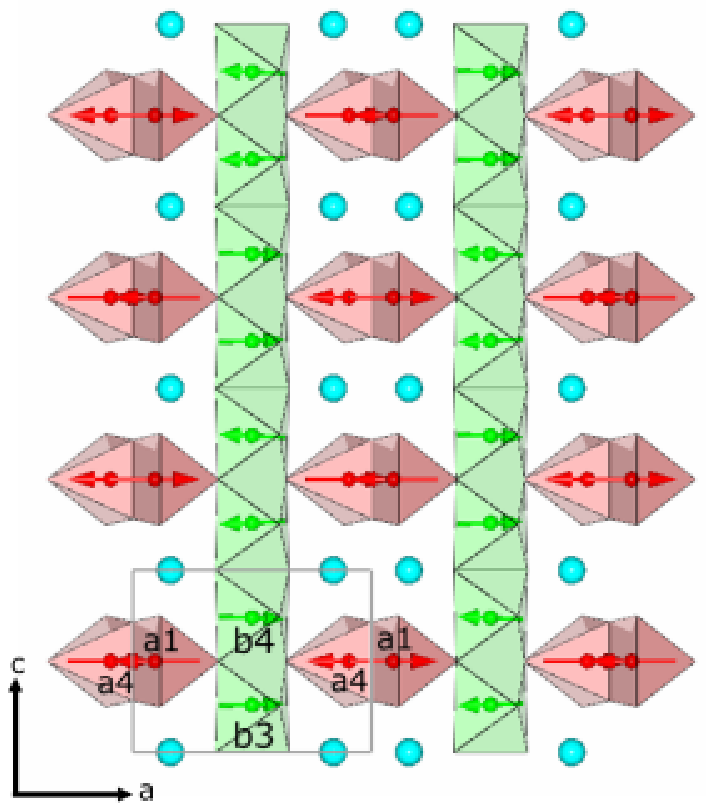}
\caption{(Color Online) Magnetic structure of $BiMn_{2}O_{5}$
projected in the \emph{ac} plane shown within four unit-cells along
\textit{c} and two unit-cells along \textit{a}. The projection is
shown separately for magnetic sites belonging to the first AFM chain
(left) and the second chain (right) (See text for details). The
green and red arrows represent magnetic moments on $Mn^{4+}$ and
$Mn^{3+}$ sites respectively. Corresponding Mn-O polyhedra are shown
with the same colors. Blue sphere represent Bi ions. The different
sites are labeled according to Table \ref{BiMn2O5table}}
\label{Bi_ac}
\end{figure*}

\BI\ is the only member of the series that shows magnetic ordering
below T=39 K with a propagation vector
$\overrightarrow{k}$=($\frac{1}{2}$, 0, $\frac{1}{2}$) and no
transitions to incommensurate magnetic order. Because of the commensurate
propagation vector representation
analysis leads to stricter constraints than those applicable to \Y\ and \HO,
they are described in detail by Munoz
et al.\cite{MunozPRB_Bi}. No phases are required at any of the sites since
$\overrightarrow{k}$ is half a reciprocal lattice vector. The magnetic representation
contains two irreducible representations of dimension 2. Only
magnetic modes belonging to the representation $\Gamma_1$ fit our
single crystal data, in agreement with the results of Munoz et al.
The magnetic moments on Mn$^{4+}$ sites are constrained in the
following way: sites b1 and b2 have the same a and b components but
opposite c components. The same relation holds for sites b3 and b4.
However, there are no relationships imposed by symmetry between the
two set of sites mentioned above, b1/b2 on one hand and b3/b4 on the
other. For the Mn$^{3+}$ sites, the moments are constrained to
lie in the \textit{ab}-plane but can have any magnitude and
orientation within it.
 In the work of Munoz et
al., the additional physical constraint of equal moments, not
imposed by symmetry, was introduced. Here, the large number of
independent magnetic reflections collected (204) allows the
simultaneous refinement of all fourteen parameters. The result of
the refinement is shown in Fig. \ref{refinements} and the
corresponding magnetic structure is displayed in Fig. \ref{Bi_ab}
and Fig. \ref{Bi_ac}. A R(F) factor of 4.5\% has been obtained with
a $\chi^{2}$ value of 18.2. \BI\ has a very similar magnetic structure
to that described previously for \Y, for instance the  antiferromagnetic
...-Mn$^{4+}$-Mn$^{3+}$-Mn$^{3+}$-Mn$^{4+}$-... chains in the
\textit{ab}-plane. This is at variance with the structure proposed  by
Munoz, in which there is a non-collinear arrangement of Mn$^{4+}$ and
Mn$^{3+}$ moments within one AFM chain. The angular deviation of the
AFM chains from the \textit{a}-axis ($\sim$8$^{\circ}$)
is less pronounced than in the \Y\ and \HO\ compounds. Also unlike
\Y\ and \HO, we found almost zero z-components for the Mn$^{4+}$
moments. The magnetic point group is m2m and the magnetic structures
of the two possible domains are related by inversion symmetry, as
previously described, resulting in flipping the central AFM chain in
Fig. \ref{Bi_ab}.

\section{Origin of ferroelectricity}

\indent In this section, we discuss the possible mechanism that will
promote a ferroelectric state in the magnetically ordered phase.
First of all it is worth noting that irrespective of the microscopic
mechanism involved, the point group symmetry m2m derived from the
magnetic structures of \Y, \HO, and \BI, is consistent with
ferroelectric order only along the \textit{b}-axis. Secondly, it is
important to discuss the possible microscopic mechanisms, in the
light of the additional weak magnetic components observed along the
\textit{c}-axis for \Y\ and \HO\ and not identified previously by
powder diffraction. This small out of plane component introduces a
modulation that resembles a cycloid, even though it does not
correspond to a homogeneous rotation from site to site, as seen in
Fig. \ref{y_ac} for example. However, this modulation could in
principle give rise to ferroelectricity based on antisymmetric
exchange interaction, a mechanism proposed in several multiferroics,
and its strength will be proportional to the spin-orbit coupling,
potentially large for octahedral Mn$^{4+}$. Obviously, the absence
of such a modulation in \BI, multiferroic with a value of the
electrical polarization comparable to \Y\ and \HO, does not support
this picture. Therefore we argue that our model based on
symmetric-exchange, and proposed initially in \cite{ChaponPRL_Tb},
is the principal mechanism driving ferroelectric order. The recent
observation of electro-magnons\cite{Sushkov_PRL07} active only for
certain polarization directions, also supports a model based on
symmetric-exchange interaction. This model is relevant since the
magnetic configuration in the \textit{ab}-plane for the three
compounds studied here is frustrated, i.e. the magnetic energy is
invariant by flipping any AFM chain in the lattice. Such a
configuration will induce atomic displacements on some of the ions
in the crystal (we speculated that mainly Mn$^{3+}$ ions are
involved based on structural anomalies related to this site) that
will lower the magnetic energy\cite{ChaponPRL_Y,
Garcia-Flores_JAP07, Cruz_JAP06}. This comes at the cost of elastic
energy, quadratic with the displacement, but for small displacements
the gain in magnetic energy (linear with displacements) overcomes
the elastic energy cost. Magneto-striction due to symmetric exchange
should be much stronger than the DM interaction, which is a
relativistic effect\cite{Yamasaki_PRL07, Lawes_PRL05,
Katsura_PRL05}. However, it is possible that this later mechanism
contributes to the total electric polarization in the \Y\ and \HO\
compounds but should be of minor importance. In fact, the two mechanisms
are likely to be strongly coupled. It is important to note that
applying the inversion symmetry operation not only flips one of the AFM
chains but also changes the rotation direction of the
cycloidal modulation. Therefore the direction of the electric
polarization in domain two is opposite to that in domain
one, whatever the mechanism considered. The observation of cycloidal
structures for many of these compounds is intriguing, and is
unlikely to be a coincidence. In general, two mechanisms are known
to generate helicoidal-type structures; competition between nearest
and next-nearest neighbor interaction and direct DM interaction,
resulting from the loss of a center of symmetry relating magnetic
ions.  It is easy to show that in none of the
magnetic structures described here does the cycloidal modulation
lower the  next-near neighbor magnetic
energy, regardless of the sign of the interaction.  This suggests
that the observed c-axis modulation may be the effect, rather than
the cause, of the loss of the center of symmetry relating Mn$^{4+}$ ions.

\section{Summary}

\indent The magnetic structures in the ferroelectric/commensurate
magnetic regime of three compounds of the family
\emph{R}Mn$_{2}$O$_{5}$ with \emph{R}= Y, Ho, and Bi, have been
determined by single crystal neutron diffraction. For \Y\ and
\HO\, the main features of the magnetic structures are in
agreement with previous models derived from powder neutron
diffraction characterised, in particular, by collinear
antiferromagnetic zig-zag chains in the \emph{ab}-plane. An
additional small component of moment  parallel to
the \emph{c}-axis, and modulated in quadrature with the major components
has been identified. This component was not detected in earlier experiments.
The data also allow a more precise determination of the directions of the
Ho magnetic moments. These results indicate the importance of super-exchange interactions in the magnetic ordering of Ho.
The magnetic structure of \BI\ is very
similar to that of the Y and Ho analogs, but with the  magnetic
moments more closely confined to the \textit{ab}-plane. It also contains
zig-zag AFM chains, in disagreement with what was previously reported\cite{MunozPRB_Bi}.
This comparative study strongly supports symmetric exchange as the
principal mechanism leading to ferroelectricity since the small
non-collinearity of the magnetic moments within chains along c is
observed only for the \Y\ and \HO\ compounds. This does not
preclude a potential, but much weaker contribution from Dzyaloshinskii-Moriya interactions.

\section{Aknowledgements}
We would like to thank B. Ouladdiaf and G. Mcintyre from the Institut Laue-Langevin (France) for useful discussions, and the valuable help during the D10 experiment and the crystal alignment with the \emph{OrientExpress} instrument.
We acknowledge partial support from the European Commission ("Construction of New Infrastructures", ISIS Target Station II, contract no. 011723).

\section{Note added in proof}
During the final elaboration of this paper, we became aware of a recent manuscript by H. Kimura \textit{et al.}, which is now in print \cite{Kimura_JPSJ07}.  Kimura \textit{et al.} investigated the crystal and magnetic structures of \Y, \HO\ and ErMn$_{2}$O$_{5}$ by single crystal neutron diffraction.  Although the data analysis methodology is different from ours (they employ a large supercell rather than Fourier components), the agreement with our results on the compounds common to the two studies is excellent.  Our work includes the refinement of the BiMn$_{2}$O$_{5}$, which, we believe, is crucial in clarifying the correlations between the different spin component and the electrical polarisation.

\newpage

\begin{figure*}[h!]
\includegraphics[scale=1.40, angle=0]{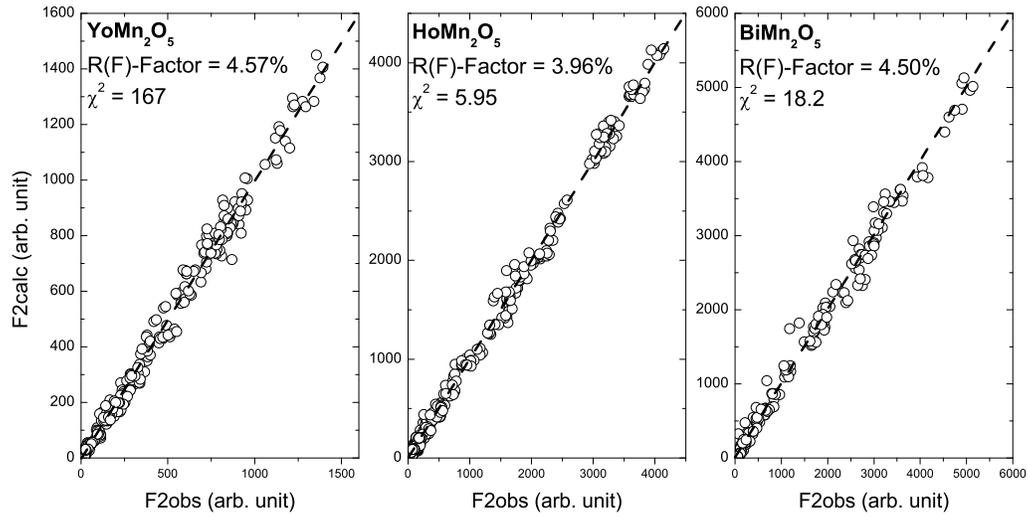}
\caption{Results of the refinement for the commensurate magnetic
structures for three compounds- $YMn_{2}O_{5}$, $HoMn_{2}O_{5}$ and
$BiMn_{2}O_{5}$. The experimental structure factors (times a scale
factor) are plotted against the calculated ones. The agreement
factors- magnetic R(F) factors and $\chi^2$- are shown. See text for
details of the magnetic models.} \label{refinements}
\label{refinements}
\end{figure*}

\newpage

\newpage

\newpage

\newpage

\newpage

\newpage

\newpage

\newpage

\newpage

\newpage
\begin{table}[h!]
\begin{center}
\begin{tabular}{c|c}
Symmetry operation & Domain generated \\ \hline
$\{1|000\}$ \ &\ D1  \\
$\{\overline{1}|000\}$ $\ast$ &\ D2 \\
$\{m_{xy0}|000\}$ $\ast$ \ &\ D1 \\
$\{2_z|000\}$ \ &\ D2\ \\
$\{m_{0yz}|0\frac{1}{2}0\}$ \ &\ D1\ \\
$\{m_{x0z}|\frac{1}{2}00\}$ &\ D2\ \\
$\{2_{x00}|\frac{1}{2}00\}$ $\ast$ \ &\ D2 \\
$\{2_{0y0}|0\frac{1}{2}0\}$  $\ast$ \ &\ D1 \\
\end{tabular}
\end{center}
\caption{Action of the symmetry operations of the paramagnetic group
on the magnetic structure of domain 1. Domain one and two are
abbreviated by D1 and D2. Symmetry operations are shown in the Seitz
notation. The $\ast$ symbol indicates combination of the symmetry
operation with complex conjugation operation.} \label{SummetryTable}
\end{table}

\newpage
\begin{table*}[h!]
\begin{center}
\begin{tabular}{lllllllccccccccc|}
Label \ & Atom & Position & $M_{x}$ & $M_{y}$ & $I_{z}$ & Ph \\ \hline
&&&&&&\\
b1 \ &\ $Mn^{4+}$\ &\ (0 0.5 0.25513)       &\ 2.01(5)  &\ -0.47(8) &\ 0.36(11)  &\ 0.095(2) & \\
b2 \ &\ $Mn^{4+}$\ &\ (0 0.5 0.74487)       &\ 2.13(5)  &\ -0.53(9) &\ 0.44(11)  &\ 0.156(2) & \\
b3 \ &\ $Mn^{4+}$\ &\ (0.5 0 0.25513)       &\ -2.07(5) &\ -0.53(8) &\ 0.4(1)    &\ 0.095(2) & \\
b4 \ &\ $Mn^{4+}$\ &\ (0.5 0 0.74487)       &\ -1.99(5) &\ -0.51(8) &\ 0.45(11)  &\ 0.156(2) & \\
a1 \ &\ $Mn^{3+}$\ &\ (0.08805 0.85079 0.5) &\ 3.16(6)  &\ 0.59(9) &\ -0.46(12)  &\ 0.125    & \\
a2 \ &\ $Mn^{3+}$\ &\ (0.41195 0.35079 0.5) &\ -3.18(6) &\ 0.55(9)  &\ -0.50(12) &\ 0.125    & \\
a3 \ &\ $Mn^{3+}$\ &\ (0.58805 0.64921 0.5) &\ 3.01(5)  &\ -0.67(9) &\ 0.56(13)  &\ 0.125    & \\
a4 \ &\ $Mn^{3+}$\ &\ (0.91195 0.14921 0.5) &\ 2.99(5)  &\ 0.65(9)  &\ -0.60(13) &\ 0.125    & \\
\end{tabular}
\end{center}
\caption{Magnetic Fourier coefficients obtained from least-square
refinements of the single crystal diffraction data for
$YMn_{2}O_{5}$ at $T=25\ K$ (See text for details). The Fourier
coefficients for each site in the primitive unit-cell are given
along crystallographic direction (x,y,z) and labeled M and I when
real and imaginary, respectively. The phase for each site (Ph.) is
given in units of 2$\pi$. Error bars are given for all parameters
within parenthesis.} \label{YMn2O5Fixedtable}
\end{table*}

\newpage
\begin{table*}[h!]
\begin{center}
\begin{tabular}{lllllllccccccccc|}
Label \ & Atom & Position & $M_{x}$ & $M_{y}$ & $I_{z}$ & Ph \\ \hline
&&&&&&\\
b1 \ &\ $Mn^{4+}$\ &\ (0 0.5 0.2558)     &\ -2.25(4)  &\ 0.58(5)   &\ -0.38(8)  &\ 0.088(2)    & \\
b2 \ &\ $Mn^{4+}$\ &\ (0 0.5 0.7442)     &\ -2.32(4)  &\ 0.59(5)   &\ -0.39(8)  &\ 0.161(2)    & \\
b3 \ &\ $Mn^{4+}$\ &\ (0.5 0 0.2558)     &\ 2.26(4)   &\ 0.59(5)   &\ -0.51(8)  &\ 0.088(2)    & \\
b4 \ &\ $Mn^{4+}$\ &\ (0.5 0 0.7442)     &\ 2.19(4)   &\ 0.55(5)   &\ -0.49(8)  &\ 0.161(2)    & \\
a1 \ &\ $Mn^{3+}$\ &\ (0.0885 0.849 0.5) &\ -3.53(4)  &\ -0.43(5)  &\ 0.58(9)   &\ 0.125         & \\
a2 \ &\ $Mn^{3+}$\ &\ (0.4115 0.349 0.5) &\ 3.49(4)   &\ -0.73(5)  &\ 0.46(9)   &\ 0.125         & \\
a3 \ &\ $Mn^{3+}$\ &\ (0.5885 0.651 0.5) &\ -3.29(4)  &\ 0.85(6)   &\ -0.60(8)  &\ 0.125         & \\
a4 \ &\ $Mn^{3+}$\ &\ (0.9115 0.151 0.5) &\ -3.27(4)  &\ -0.72(6)  &\ 0.65(8)   &\ 0.125         & \\
c1 \ &\ $Ho^{3+}$\ &\ (0.1392 0.1713 0)  &\ 0.03(3)   &\ -1.09(4)  &\ 0.18(7)   &\ -0.001(2)     & \\
c2 \ &\ $Ho^{3+}$\ &\ (0.3608 0.6713 0)  &\ -0.03(3)  &\ -1.14(4)  &\ 0.16(6)   &\ -0.001(2)     & \\
c3 \ &\ $Ho^{3+}$\ &\ (0.6392 0.3284 0)  &\ -1.27(4)  &\ -0.04(4)  &\ -0.16(6)  &\ -0.001(2)     & \\
c4 \ &\ $Ho^{3+}$\ &\ (0.8606 0.8287 0)  &\ -1.25(4)  &\ -0.08(4)  &\ 0.24(6)   &\ -0.001(2)     & \\
\end{tabular}
\end{center}
\caption{Magnetic Fourier coefficients obtained from least-square
refinements of the single crystal diffraction data for
$HoMn_{2}O_{5}$ at $T=25\ K$ (See text for details). The Fourier
coefficients for each site in the primitive unit-cell are given
along crystallographic direction (x,y,z) and labeled M and I when
real and imaginary, respectively. The phase for each site (Ph.) is
given in units of 2$\pi$. Error bars are given for all parameters
within parenthesis.}
\label{HoMn2O5Fixedtable}
\end{table*}

\newpage
\begin{table*}[h!]
\begin{center}
\begin{tabular}{llllllccccccccc|}
Label \ & Atom & Position & $M_{x}$ & $M_{y}$ & $M_{z}$ & Ph \\
\hline
&&&&&&\\
b1 \ &\ $Mn^{4+}$\ &\ (0 0.5 0.2613)         &\   2.10(3) &\  -0.33(6)  &\  0.25(6) &\ 0.0 & \\
b2 \ &\ $Mn^{4+}$\ &\ (0 0.5 0.7387)         &\   2.10(3) &\  -0.33(6)  &\ -0.25(6) &\ 0.0 & \\
b3 \ &\ $Mn^{4+}$\ &\ (0.5 0 0.2613)         &\   2.07(3) &\   0.56(6)  &\  0.08(6) &\ 0.0 & \\
b4 \ &\ $Mn^{4+}$\ &\ (0.5 0 0.7387)         &\   2.07(3) &\   0.56(6)  &\ -0.08(6) &\ 0.0 & \\
a1 \ &\ $Mn^{3+}$\ &\ (0.0926 0.8516 0.5)    &\  -2.83(5) &\  -0.23(10) &\  0.000    &\ 0.0 & \\
a2 \ &\ $Mn^{3+}$\ &\ (0.4074 0.3516 0.5)    &\  -2.83(5) &\   0.33(10) &\  0.000   &\ 0.0 & \\
a3 \ &\ $Mn^{3+}$\ &\ (0.5926 0.6484 0.5)    &\   2.80(5) &\  -0.34(9)  &\  0.000   &\ 0.0 & \\
a4 \ &\ $Mn^{3+}$\ &\ (0.9074 0.1484 0.5)    &\  -2.74(5) &\  -0.64(10) &\  0.000   &\ 0.0 & \\
\end{tabular}
\end{center}
\caption{Magnetic Fourier coefficients obtained from least-square
refinements of the single crystal diffraction data for
$BiMn_{2}O_{5}$ at $T=10\ K$ (See text for details). The Fourier
coefficients for each site (M, real) in the primitive unit-cell are given
along crystallographic direction (x,y,z). The phase for each site (Ph.) is
given in units of 2$\pi$. Error bars are given for all parameters
within parenthesis.}
\label{BiMn2O5table}
\end{table*}

\newpage

\begin{thebibliography}{31}
\expandafter\ifx\csname natexlab\endcsname\relax\def\natexlab#1{#1}\fi
\expandafter\ifx\csname bibnamefont\endcsname\relax
  \def\bibnamefont#1{#1}\fi
\expandafter\ifx\csname bibfnamefont\endcsname\relax
  \def\bibfnamefont#1{#1}\fi
\expandafter\ifx\csname citenamefont\endcsname\relax
  \def\citenamefont#1{#1}\fi
\expandafter\ifx\csname url\endcsname\relax
  \def\url#1{\texttt{#1}}\fi
\expandafter\ifx\csname urlprefix\endcsname\relax\def\urlprefix{URL }\fi
\providecommand{\bibinfo}[2]{#2}
\providecommand{\eprint}[2][]{\url{#2}}

\bibitem[{\citenamefont{Kimura et~al.}(2003)\citenamefont{Kimura, Goto,
  Shintani, Ishizaka, Arima, and Tokura}}]{Kimura_Nature03}
\bibinfo{author}{\bibfnamefont{T.}~\bibnamefont{Kimura}},
  \bibinfo{author}{\bibfnamefont{T.}~\bibnamefont{Goto}},
  \bibinfo{author}{\bibfnamefont{H.}~\bibnamefont{Shintani}},
  \bibinfo{author}{\bibfnamefont{K.}~\bibnamefont{Ishizaka}},
  \bibinfo{author}{\bibfnamefont{T.}~\bibnamefont{Arima}}, \bibnamefont{and}
  \bibinfo{author}{\bibfnamefont{Y.}~\bibnamefont{Tokura}},
  \bibinfo{journal}{Nature} \textbf{\bibinfo{volume}{426}}, \bibinfo{pages}{55}
  (\bibinfo{year}{2003}).

\bibitem[{\citenamefont{Hur et~al.}(2004)\citenamefont{Hur, Park, Sharma, Ahn,
  Guha, and Cheong}}]{Hur_Nature04}
\bibinfo{author}{\bibfnamefont{N.}~\bibnamefont{Hur}},
  \bibinfo{author}{\bibfnamefont{S.}~\bibnamefont{Park}},
  \bibinfo{author}{\bibfnamefont{P.~A.} \bibnamefont{Sharma}},
  \bibinfo{author}{\bibfnamefont{J.~S.} \bibnamefont{Ahn}},
  \bibinfo{author}{\bibfnamefont{S.}~\bibnamefont{Guha}}, \bibnamefont{and}
  \bibinfo{author}{\bibfnamefont{S.-W.} \bibnamefont{Cheong}},
  \bibinfo{journal}{Nature} \textbf{\bibinfo{volume}{429}},
  \bibinfo{pages}{392} (\bibinfo{year}{2004}).

\bibitem[{\citenamefont{Cheong and Mostovoy}(2007)}]{Cheong_NatureMat}
\bibinfo{author}{\bibfnamefont{S.-W.} \bibnamefont{Cheong}} \bibnamefont{and}
  \bibinfo{author}{\bibfnamefont{M.}~\bibnamefont{Mostovoy}},
  \bibinfo{journal}{Nature Mater.} \textbf{\bibinfo{volume}{6}},
  \bibinfo{pages}{13} (\bibinfo{year}{2007}).

\bibitem[{\citenamefont{Kagomiya et~al.}(2003)\citenamefont{Kagomiya,
  Matsumoto, Khon, Fukuda, Shoubu, Kimura, Noda, and Ikeda}}]{ferroelectrics}
\bibinfo{author}{\bibfnamefont{I.}~\bibnamefont{Kagomiya}},
  \bibinfo{author}{\bibfnamefont{S.}~\bibnamefont{Matsumoto}},
  \bibinfo{author}{\bibfnamefont{K.}~\bibnamefont{Khon}},
  \bibinfo{author}{\bibfnamefont{Y.}~\bibnamefont{Fukuda}},
  \bibinfo{author}{\bibfnamefont{T.}~\bibnamefont{Shoubu}},
  \bibinfo{author}{\bibfnamefont{H.}~\bibnamefont{Kimura}},
  \bibinfo{author}{\bibfnamefont{Y.}~\bibnamefont{Noda}}, \bibnamefont{and}
  \bibinfo{author}{\bibfnamefont{N.}~\bibnamefont{Ikeda}},
  \bibinfo{journal}{Ferroelectrics} \textbf{\bibinfo{volume}{286}},
  \bibinfo{pages}{889} (\bibinfo{year}{2003}).

\bibitem[{\citenamefont{Eerenstein et~al.}(2006)\citenamefont{Eerenstein,
  Mathur, and Scott}}]{Eerenstein_Nature}
\bibinfo{author}{\bibfnamefont{W.}~\bibnamefont{Eerenstein}},
  \bibinfo{author}{\bibfnamefont{N.~D.} \bibnamefont{Mathur}},
  \bibnamefont{and} \bibinfo{author}{\bibfnamefont{J.~F.} \bibnamefont{Scott}},
  \bibinfo{journal}{Nature} \textbf{\bibinfo{volume}{442}},
  \bibinfo{pages}{759} (\bibinfo{year}{2006}).

\bibitem[{\citenamefont{García-Flores et~al.}(2007)\citenamefont{García-Flores,
  Granado, Martinho, Rettori, Golovenchits, Sanina, Oseroff, Park, and
  Cheong}}]{Garcia-Flores_JAP07}
\bibinfo{author}{\bibfnamefont{A.~F.} \bibnamefont{García-Flores}},
  \bibinfo{author}{\bibfnamefont{E.}~\bibnamefont{Granado}},
  \bibinfo{author}{\bibfnamefont{H.}~\bibnamefont{Martinho}},
  \bibinfo{author}{\bibfnamefont{C.}~\bibnamefont{Rettori}},
  \bibinfo{author}{\bibfnamefont{E.~I.} \bibnamefont{Golovenchits}},
  \bibinfo{author}{\bibfnamefont{V.~A.} \bibnamefont{Sanina}},
  \bibinfo{author}{\bibfnamefont{S.~B.} \bibnamefont{Oseroff}},
  \bibinfo{author}{\bibfnamefont{S.}~\bibnamefont{Park}}, \bibnamefont{and}
  \bibinfo{author}{\bibfnamefont{S.-W.} \bibnamefont{Cheong}},
  \bibinfo{journal}{J. Appl. Phys.} \textbf{\bibinfo{volume}{101}},
  \bibinfo{pages}{09M106} (\bibinfo{year}{2007}).

\bibitem[{\citenamefont{Aguilar et~al.}(2006)\citenamefont{Aguilar, Sushkov,
  Park, Cheong, and Drew}}]{Aguilar_PRB06}
\bibinfo{author}{\bibfnamefont{R.~V.} \bibnamefont{Aguilar}},
  \bibinfo{author}{\bibfnamefont{A.~B.} \bibnamefont{Sushkov}},
  \bibinfo{author}{\bibfnamefont{S.}~\bibnamefont{Park}},
  \bibinfo{author}{\bibfnamefont{S.-W.} \bibnamefont{Cheong}},
  \bibnamefont{and} \bibinfo{author}{\bibfnamefont{H.~D.} \bibnamefont{Drew}},
  \bibinfo{journal}{Phys. Rev. B} \textbf{\bibinfo{volume}{74}},
  \bibinfo{pages}{184404} (\bibinfo{year}{2006}).

\bibitem[{\citenamefont{Noda et~al.}(2006)\citenamefont{Noda, Kimura, Kamada,
  Osawa, Fukuda, Ishikawa, Kobayashi, Wakabayashi, Sawa, Ikeda
  et~al.}}]{Noda_PhysB06}
\bibinfo{author}{\bibfnamefont{Y.}~\bibnamefont{Noda}},
  \bibinfo{author}{\bibfnamefont{H.}~\bibnamefont{Kimura}},
  \bibinfo{author}{\bibfnamefont{Y.}~\bibnamefont{Kamada}},
  \bibinfo{author}{\bibfnamefont{T.}~\bibnamefont{Osawa}},
  \bibinfo{author}{\bibfnamefont{Y.}~\bibnamefont{Fukuda}},
  \bibinfo{author}{\bibfnamefont{Y.}~\bibnamefont{Ishikawa}},
  \bibinfo{author}{\bibfnamefont{S.}~\bibnamefont{Kobayashi}},
  \bibinfo{author}{\bibfnamefont{Y.}~\bibnamefont{Wakabayashi}},
  \bibinfo{author}{\bibfnamefont{H.}~\bibnamefont{Sawa}},
  \bibinfo{author}{\bibfnamefont{N.}~\bibnamefont{Ikeda}},
  \bibnamefont{et~al.}, \bibinfo{journal}{Phys. B}
  \textbf{\bibinfo{volume}{385}}, \bibinfo{pages}{119} (\bibinfo{year}{2006}).

\bibitem[{\citenamefont{Han and Lin}(2006)}]{Han_JAP06}
\bibinfo{author}{\bibfnamefont{T.-C.} \bibnamefont{Han}} \bibnamefont{and}
  \bibinfo{author}{\bibfnamefont{J.~G.} \bibnamefont{Lin}},
  \bibinfo{journal}{J. Appl. Phys.} \textbf{\bibinfo{volume}{99}},
  \bibinfo{pages}{08J508} (\bibinfo{year}{2006}).

\bibitem[{\citenamefont{Lin et~al.}(2005)\citenamefont{Lin, Han, and
  Chen}}]{Lin_IEEE05}
\bibinfo{author}{\bibfnamefont{J.~G.} \bibnamefont{Lin}},
  \bibinfo{author}{\bibfnamefont{T.-C.} \bibnamefont{Han}}, \bibnamefont{and}
  \bibinfo{author}{\bibfnamefont{C.-H.} \bibnamefont{Chen}},
  \bibinfo{journal}{IEEE Trans. Magn.} \textbf{\bibinfo{volume}{41}},
  \bibinfo{pages}{3440} (\bibinfo{year}{2005}).

\bibitem[{\citenamefont{Prokhnenko et~al.}(2007)\citenamefont{Prokhnenko,
  Feyerherm, Dudzik, Landsgesell, Aliouane, Chapon, and
  Argyriou}}]{Prokhnenko_PRL07}
\bibinfo{author}{\bibfnamefont{O.}~\bibnamefont{Prokhnenko}},
  \bibinfo{author}{\bibfnamefont{R.}~\bibnamefont{Feyerherm}},
  \bibinfo{author}{\bibfnamefont{E.}~\bibnamefont{Dudzik}},
  \bibinfo{author}{\bibfnamefont{S.}~\bibnamefont{Landsgesell}},
  \bibinfo{author}{\bibfnamefont{N.}~\bibnamefont{Aliouane}},
  \bibinfo{author}{\bibfnamefont{L.~C.} \bibnamefont{Chapon}},
  \bibnamefont{and} \bibinfo{author}{\bibfnamefont{D.~N.}
  \bibnamefont{Argyriou}}, \bibinfo{journal}{Phys. Rev. Lett.}
  \textbf{\bibinfo{volume}{98}}, \bibinfo{pages}{057206}
  (\bibinfo{year}{2007}).

\bibitem[{\citenamefont{Koo et~al.}(2007)\citenamefont{Koo, Song, Ji, Lee,
  Park, Jang, Yang, Park, Jeong, Lee et~al.}}]{koo_2007}
\bibinfo{author}{\bibfnamefont{J.}~\bibnamefont{Koo}},
  \bibinfo{author}{\bibfnamefont{C.}~\bibnamefont{Song}},
  \bibinfo{author}{\bibfnamefont{S.}~\bibnamefont{Ji}},
  \bibinfo{author}{\bibfnamefont{J.~S.} \bibnamefont{Lee}},
  \bibinfo{author}{\bibfnamefont{J.}~\bibnamefont{Park}},
  \bibinfo{author}{\bibfnamefont{T.~H.} \bibnamefont{Jang}},
  \bibinfo{author}{\bibfnamefont{C.~H.} \bibnamefont{Yang}},
  \bibinfo{author}{\bibfnamefont{J.~H.} \bibnamefont{Park}},
  \bibinfo{author}{\bibfnamefont{Y.~H.} \bibnamefont{Jeong}},
  \bibinfo{author}{\bibfnamefont{K.~B.} \bibnamefont{Lee}},
  \bibnamefont{et~al.}, \emph{\bibinfo{title}{Non-resonant and resonant x-ray
  scattering studies on multiferroic tbmn2o5}} (\bibinfo{year}{2007}),
  \urlprefix\url{http://www.citebase.org/abstract?id=oai:arXiv.org:0704.0533}.

\bibitem[{\citenamefont{Khomskii}(2006)}]{Khomskii_JMMM2006}
\bibinfo{author}{\bibfnamefont{D.~I.} \bibnamefont{Khomskii}},
  \bibinfo{journal}{J. Magn. Magn. Mater} \textbf{\bibinfo{volume}{306}},
  \bibinfo{pages}{1} (\bibinfo{year}{2006}).

\bibitem[{\citenamefont{Yamasaki et~al.}(2007)\citenamefont{Yamasaki, Sagayama,
  Goto, Matsuura, Hirota, Arima, and Tokura}}]{Yamasaki_PRL07}
\bibinfo{author}{\bibfnamefont{Y.}~\bibnamefont{Yamasaki}},
  \bibinfo{author}{\bibfnamefont{H.}~\bibnamefont{Sagayama}},
  \bibinfo{author}{\bibfnamefont{T.}~\bibnamefont{Goto}},
  \bibinfo{author}{\bibfnamefont{M.}~\bibnamefont{Matsuura}},
  \bibinfo{author}{\bibfnamefont{K.}~\bibnamefont{Hirota}},
  \bibinfo{author}{\bibfnamefont{T.}~\bibnamefont{Arima}}, \bibnamefont{and}
  \bibinfo{author}{\bibfnamefont{Y.}~\bibnamefont{Tokura}},
  \bibinfo{journal}{Phys. Rev. Lett.} \textbf{\bibinfo{volume}{98}},
  \bibinfo{pages}{147204} (\bibinfo{year}{2007}).

\bibitem[{\citenamefont{Lawes et~al.}(2005)\citenamefont{Lawes, Harris, Kimura,
  Rogado, Cava, Aharony, Entin-Wohlman, Yildirim, Kenzelmann, Broholm
  et~al.}}]{Lawes_PRL05}
\bibinfo{author}{\bibfnamefont{G.}~\bibnamefont{Lawes}},
  \bibinfo{author}{\bibfnamefont{A.~B.} \bibnamefont{Harris}},
  \bibinfo{author}{\bibfnamefont{T.}~\bibnamefont{Kimura}},
  \bibinfo{author}{\bibfnamefont{N.}~\bibnamefont{Rogado}},
  \bibinfo{author}{\bibfnamefont{R.~J.} \bibnamefont{Cava}},
  \bibinfo{author}{\bibfnamefont{A.}~\bibnamefont{Aharony}},
  \bibinfo{author}{\bibfnamefont{O.}~\bibnamefont{Entin-Wohlman}},
  \bibinfo{author}{\bibfnamefont{T.}~\bibnamefont{Yildirim}},
  \bibinfo{author}{\bibfnamefont{M.}~\bibnamefont{Kenzelmann}},
  \bibinfo{author}{\bibfnamefont{C.}~\bibnamefont{Broholm}},
  \bibnamefont{et~al.}, \bibinfo{journal}{Phys. Rev. Lett.}
  \textbf{\bibinfo{volume}{95}}, \bibinfo{pages}{087205}
  (\bibinfo{year}{2005}).

\bibitem[{\citenamefont{Chapon et~al.}(2006)\citenamefont{Chapon, Radaelli,
  Blake, Park, and Cheong}}]{ChaponPRL_Y}
\bibinfo{author}{\bibfnamefont{L.~C.} \bibnamefont{Chapon}},
  \bibinfo{author}{\bibfnamefont{P.~G.} \bibnamefont{Radaelli}},
  \bibinfo{author}{\bibfnamefont{G.~R.} \bibnamefont{Blake}},
  \bibinfo{author}{\bibfnamefont{S.}~\bibnamefont{Park}}, \bibnamefont{and}
  \bibinfo{author}{\bibfnamefont{S.-W.} \bibnamefont{Cheong}},
  \bibinfo{journal}{Phys. Rev. Lett.} \textbf{\bibinfo{volume}{96}},
  \bibinfo{pages}{097601} (\bibinfo{year}{2006}).

\bibitem[{\citenamefont{Sergienko et~al.}(2006)\citenamefont{Sergienko, Sen,
  and Dagotto}}]{Sergienko_PRL06}
\bibinfo{author}{\bibfnamefont{I.~A.} \bibnamefont{Sergienko}},
  \bibinfo{author}{\bibfnamefont{C.}~\bibnamefont{Sen}}, \bibnamefont{and}
  \bibinfo{author}{\bibfnamefont{E.}~\bibnamefont{Dagotto}},
  \bibinfo{journal}{Phys. Rev. Lett.} \textbf{\bibinfo{volume}{97}},
  \bibinfo{pages}{227204} (\bibinfo{year}{2006}).

\bibitem[{\citenamefont{Wilkinson et~al.}(1981)\citenamefont{Wilkinson,
  Sinclair, Gardner, Forsyth, and Wanklyn}}]{Wilkinson_JPC1981}
\bibinfo{author}{\bibfnamefont{C.}~\bibnamefont{Wilkinson}},
  \bibinfo{author}{\bibfnamefont{F.}~\bibnamefont{Sinclair}},
  \bibinfo{author}{\bibfnamefont{P.}~\bibnamefont{Gardner}},
  \bibinfo{author}{\bibfnamefont{J.~B.} \bibnamefont{Forsyth}},
  \bibnamefont{and} \bibinfo{author}{\bibfnamefont{B.~M.~R.}
  \bibnamefont{Wanklyn}}, \bibinfo{journal}{J. Phys. C: Solid State Phys.}
  \textbf{\bibinfo{volume}{14}}, \bibinfo{pages}{1671} (\bibinfo{year}{1981}).

\bibitem[{\citenamefont{Gardner et~al.}(1988)\citenamefont{Gardner, Wilkinson,
  Forsyth, and Wanklyn}}]{Gardner_JPC1988}
\bibinfo{author}{\bibfnamefont{P.~P.} \bibnamefont{Gardner}},
  \bibinfo{author}{\bibfnamefont{C.}~\bibnamefont{Wilkinson}},
  \bibinfo{author}{\bibfnamefont{J.~B.} \bibnamefont{Forsyth}},
  \bibnamefont{and} \bibinfo{author}{\bibfnamefont{B.~M.}
  \bibnamefont{Wanklyn}}, \bibinfo{journal}{J. Phys. C: Solid State Phys.}
  \textbf{\bibinfo{volume}{21}}, \bibinfo{pages}{5653} (\bibinfo{year}{1988}).

\bibitem[{\citenamefont{II et~al.}(2005)\citenamefont{II, Kiryukhin,
  Kenzelmann, Lee, Erwin, Schefer, Hur, Park, and Cheong}}]{Ratcliff_PRB05}
\bibinfo{author}{\bibfnamefont{W.~R.} \bibnamefont{II}},
  \bibinfo{author}{\bibfnamefont{V.}~\bibnamefont{Kiryukhin}},
  \bibinfo{author}{\bibfnamefont{M.}~\bibnamefont{Kenzelmann}},
  \bibinfo{author}{\bibfnamefont{S.-H.} \bibnamefont{Lee}},
  \bibinfo{author}{\bibfnamefont{R.}~\bibnamefont{Erwin}},
  \bibinfo{author}{\bibfnamefont{J.}~\bibnamefont{Schefer}},
  \bibinfo{author}{\bibfnamefont{N.}~\bibnamefont{Hur}},
  \bibinfo{author}{\bibfnamefont{S.}~\bibnamefont{Park}}, \bibnamefont{and}
  \bibinfo{author}{\bibfnamefont{S.-W.} \bibnamefont{Cheong}},
  \bibinfo{journal}{Phys. Rev. B} \textbf{\bibinfo{volume}{72}},
  \bibinfo{pages}{0600407} (\bibinfo{year}{2005}).

\bibitem[{\citenamefont{Mu$\tilde{n}$oz
  et~al.}(2002)\citenamefont{Mu$\tilde{n}$oz, Alonso, Casais,
  Mart\'{i}nez-Lope, Mart\'{i}nez, and Fern\'{a}ndez-D\'{i}az}}]{MunozPRB_Bi}
\bibinfo{author}{\bibfnamefont{A.}~\bibnamefont{Mu$\tilde{n}$oz}},
  \bibinfo{author}{\bibfnamefont{J.~A.} \bibnamefont{Alonso}},
  \bibinfo{author}{\bibfnamefont{M.~T.} \bibnamefont{Casais}},
  \bibinfo{author}{\bibfnamefont{M.~J.} \bibnamefont{Mart\'{i}nez-Lope}},
  \bibinfo{author}{\bibfnamefont{J.~L.} \bibnamefont{Mart\'{i}nez}},
  \bibnamefont{and}
  \bibinfo{author}{\bibfnamefont{M.}~\bibnamefont{Fern\'{a}ndez-D\'{i}az}},
  \bibinfo{journal}{Phys. Rev. B} \textbf{\bibinfo{volume}{65}},
  \bibinfo{pages}{144423} (\bibinfo{year}{2002}).

\bibitem[{\citenamefont{Rodriguez-Carvajal}(1993)}]{Fullprof}
\bibinfo{author}{\bibfnamefont{J.}~\bibnamefont{Rodriguez-Carvajal}},
  \bibinfo{journal}{Physica\ B} \textbf{\bibinfo{volume}{192}},
  \bibinfo{pages}{55} (\bibinfo{year}{1993}).

\bibitem[{\citenamefont{Blake et~al.}(2005)\citenamefont{Blake, Chapon,
  Radaelli, Park, Hur, Cheong, and Rodriguez-Carvajal}}]{BlakePRB}
\bibinfo{author}{\bibfnamefont{G.~R.} \bibnamefont{Blake}},
  \bibinfo{author}{\bibfnamefont{L.~C.} \bibnamefont{Chapon}},
  \bibinfo{author}{\bibfnamefont{P.~G.} \bibnamefont{Radaelli}},
  \bibinfo{author}{\bibfnamefont{S.}~\bibnamefont{Park}},
  \bibinfo{author}{\bibfnamefont{N.}~\bibnamefont{Hur}},
  \bibinfo{author}{\bibfnamefont{S.-W.} \bibnamefont{Cheong}},
  \bibnamefont{and}
  \bibinfo{author}{\bibfnamefont{J.}~\bibnamefont{Rodriguez-Carvajal}},
  \bibinfo{journal}{Phys. Rev.\ B} \textbf{\bibinfo{volume}{71}},
  \bibinfo{pages}{214402} (\bibinfo{year}{2005}).

\bibitem[{\citenamefont{Radaelli and Chapon}(2007)}]{radaelli-2006}
\bibinfo{author}{\bibfnamefont{P.~G.} \bibnamefont{Radaelli}} \bibnamefont{and}
  \bibinfo{author}{\bibfnamefont{L.~C.} \bibnamefont{Chapon}},
  \bibinfo{journal}{Phys. Rev. B} \textbf{\bibinfo{volume}{76}},
  \bibinfo{pages}{054428} (\bibinfo{year}{2007}).

\bibitem[{\citenamefont{Pannetier et~al.}(1990)\citenamefont{Pannetier,
  Bassas-Alsina, Rodriguez-Carvajal, and Caignaert}}]{Juan_Nature}
\bibinfo{author}{\bibfnamefont{J.}~\bibnamefont{Pannetier}},
  \bibinfo{author}{\bibfnamefont{J.}~\bibnamefont{Bassas-Alsina}},
  \bibinfo{author}{\bibfnamefont{J.}~\bibnamefont{Rodriguez-Carvajal}},
  \bibnamefont{and}
  \bibinfo{author}{\bibfnamefont{V.}~\bibnamefont{Caignaert}},
  \bibinfo{journal}{Nature} \textbf{\bibinfo{volume}{346}},
  \bibinfo{pages}{343} (\bibinfo{year}{1990}).

\bibitem[{\citenamefont{Amoretti et~al.}(1994)\citenamefont{Amoretti, Caciuffo,
  Santini, Francescangeli, Goremychkin, Osborn, Calestani, Sparpaglione, and
  Bonoldi}}]{Amoretti_CF94}
\bibinfo{author}{\bibfnamefont{G.}~\bibnamefont{Amoretti}},
  \bibinfo{author}{\bibfnamefont{R.}~\bibnamefont{Caciuffo}},
  \bibinfo{author}{\bibnamefont{Santini}},
  \bibinfo{author}{\bibfnamefont{O.}~\bibnamefont{Francescangeli}},
  \bibinfo{author}{\bibfnamefont{E.}~\bibnamefont{Goremychkin}},
  \bibinfo{author}{\bibfnamefont{R.}~\bibnamefont{Osborn}},
  \bibinfo{author}{\bibfnamefont{G.}~\bibnamefont{Calestani}},
  \bibinfo{author}{\bibfnamefont{M.}~\bibnamefont{Sparpaglione}},
  \bibnamefont{and} \bibinfo{author}{\bibfnamefont{L.}~\bibnamefont{Bonoldi}},
  \bibinfo{journal}{Physica C} \textbf{\bibinfo{volume}{221}},
  \bibinfo{pages}{227} (\bibinfo{year}{1994}).

\bibitem[{\citenamefont{Chapon et~al.}(2004)\citenamefont{Chapon, Blake,
  Gutmann, Park, Hur, Radaelli, and Cheong}}]{ChaponPRL_Tb}
\bibinfo{author}{\bibfnamefont{L.~C.} \bibnamefont{Chapon}},
  \bibinfo{author}{\bibfnamefont{G.~R.} \bibnamefont{Blake}},
  \bibinfo{author}{\bibfnamefont{M.~J.} \bibnamefont{Gutmann}},
  \bibinfo{author}{\bibfnamefont{S.}~\bibnamefont{Park}},
  \bibinfo{author}{\bibfnamefont{N.}~\bibnamefont{Hur}},
  \bibinfo{author}{\bibfnamefont{P.~G.} \bibnamefont{Radaelli}},
  \bibnamefont{and} \bibinfo{author}{\bibfnamefont{S.-W.}
  \bibnamefont{Cheong}}, \bibinfo{journal}{Phys. Rev. Lett.}
  \textbf{\bibinfo{volume}{93}}, \bibinfo{pages}{177402}
  (\bibinfo{year}{2004}).

\bibitem[{\citenamefont{Sushkov et~al.}(2007)\citenamefont{Sushkov, Aguilar,
  Park, heon, and Drew}}]{Sushkov_PRL07}
\bibinfo{author}{\bibfnamefont{A.~B.} \bibnamefont{Sushkov}},
  \bibinfo{author}{\bibfnamefont{R.~V.} \bibnamefont{Aguilar}},
  \bibinfo{author}{\bibfnamefont{S.}~\bibnamefont{Park}},
  \bibinfo{author}{\bibfnamefont{S.-W.} \bibnamefont{heon}}, \bibnamefont{and}
  \bibinfo{author}{\bibfnamefont{H.~D.} \bibnamefont{Drew}},
  \bibinfo{journal}{Phys. Rev. Lett.} \textbf{\bibinfo{volume}{98}},
  \bibinfo{pages}{027202} (\bibinfo{year}{2007}).

\bibitem[{\citenamefont{dela Cruz et~al.}(2006)\citenamefont{dela Cruz, Yen,
  Lorenz, Park, Cheong, Gospodinov, Ratcliff, Lynn, and Chu}}]{Cruz_JAP06}
\bibinfo{author}{\bibfnamefont{C.~R.} \bibnamefont{dela Cruz}},
  \bibinfo{author}{\bibfnamefont{F.}~\bibnamefont{Yen}},
  \bibinfo{author}{\bibfnamefont{B.}~\bibnamefont{Lorenz}},
  \bibinfo{author}{\bibfnamefont{S.}~\bibnamefont{Park}},
  \bibinfo{author}{\bibfnamefont{S.-W.} \bibnamefont{Cheong}},
  \bibinfo{author}{\bibfnamefont{M.~M.} \bibnamefont{Gospodinov}},
  \bibinfo{author}{\bibfnamefont{W.}~\bibnamefont{Ratcliff}},
  \bibinfo{author}{\bibfnamefont{J.~W.} \bibnamefont{Lynn}}, \bibnamefont{and}
  \bibinfo{author}{\bibfnamefont{C.~W.} \bibnamefont{Chu}},
  \bibinfo{journal}{J. Appl. Phys.} \textbf{\bibinfo{volume}{99}},
  \bibinfo{pages}{08R103} (\bibinfo{year}{2006}).

\bibitem[{\citenamefont{Katsura et~al.}(2005)\citenamefont{Katsura, Nagaosa,
  and Balatsky}}]{Katsura_PRL05}
\bibinfo{author}{\bibfnamefont{H.}~\bibnamefont{Katsura}},
  \bibinfo{author}{\bibfnamefont{N.}~\bibnamefont{Nagaosa}}, \bibnamefont{and}
  \bibinfo{author}{\bibfnamefont{A.~V.} \bibnamefont{Balatsky}},
  \bibinfo{journal}{Phys. Rev. Lett.} \textbf{\bibinfo{volume}{95}},
  \bibinfo{pages}{057205} (\bibinfo{year}{2005}).

\bibitem[{\citenamefont{Kimura et~al.}(2007)\citenamefont{Kimura, Kobayashi,
  Fukuda, Osawa, Kamada, Noda, Kagomiya, and Kohn}}]{Kimura_JPSJ07}
\bibinfo{author}{\bibfnamefont{H.}~\bibnamefont{Kimura}},
  \bibinfo{author}{\bibfnamefont{S.}~\bibnamefont{Kobayashi}},
  \bibinfo{author}{\bibfnamefont{Y.}~\bibnamefont{Fukuda}},
  \bibinfo{author}{\bibfnamefont{T.}~\bibnamefont{Osawa}},
  \bibinfo{author}{\bibfnamefont{Y.}~\bibnamefont{Kamada}},
  \bibinfo{author}{\bibfnamefont{Y.}~\bibnamefont{Noda}},
  \bibinfo{author}{\bibfnamefont{I.}~\bibnamefont{Kagomiya}}, \bibnamefont{and}
  \bibinfo{author}{\bibfnamefont{K.}~\bibnamefont{Kohn}}, \bibinfo{journal}{J.
  Phys. Society Japan} \textbf{\bibinfo{volume}{76}}, \bibinfo{pages}{074706}
  (\bibinfo{year}{2007}).

\end{thebibliography}

\end{document}